\documentclass[pdftex,twocolumn,10pt,letterpaper]{extarticle}

\newcommand{\system}{PipeDream\xspace}

\newcommand{\AUTHORS}{Amar Phanishayee}
\newcommand{\TITLE}{\system{}: Fast and Efficient Pipeline Parallel DNN Training \vspace{-0.22in}}
\newcommand{\KEYWORDS}{}
\newcommand{\CONFERENCE}{}
\newcommand{\PAGENUMBERS}{yes}       
\newcommand{\COLOR}{yes}
\newcommand{\showComments}{no}
\newcommand{\comment}[1]{}
\newcommand{\onlyAbstract}{no}
\newcommand{\showDraftMark}{no}

\usepackage[T1]{fontenc}
\usepackage[utf8]{inputenc}
\usepackage{newtxtext,newtxmath}       
\usepackage{bm}                        
\usepackage{inconsolata}              
\usepackage[scaled=0.92]{helvet}
\usepackage{booktabs}
\usepackage{subcaption}

\usepackage{ifthen}
\usepackage{comment}

\ifthenelse{\equal{\PAGENUMBERS}{yes}}{%
\usepackage[nohead,
            left=1in,right=1in,top=1in,
            footskip=0.5in,bottom=1in,     
            columnsep=0.25in
            ]{geometry}
}{%
\usepackage[noheadfoot,left=1in,right=1in,top=1in,
            footskip=0.5in,bottom=1in,
            columnsep=0.25in
	    ]{geometry}
}

\usepackage[font={small}, labelfont=bf]{caption}

\usepackage{titlesec}

\usepackage{enumitem}
\setlist{itemsep=0pt,parsep=0pt}             

\usepackage{fancyhdr}

\ifthenelse{\equal{\PAGENUMBERS}{yes}}{%
  \pagestyle{plain}
}{%
  \pagestyle{empty}
}

\ifthenelse{\equal{\showDraftMark}{yes}}{%
  \usepackage{draftfoot}
}{}

\usepackage[numbers]{natbib}

\usepackage{booktabs}
\usepackage{color}
\usepackage{xcolor}
\usepackage{colortbl}
\usepackage{float}                           
\usepackage{titling}                        

\ifthenelse{\equal{\COLOR}{yes}}{%
  \usepackage[colorlinks,citecolor=blue]{hyperref}
}{%
  \usepackage[pdfborder={0 0 0}]{hyperref}
}
\usepackage{url}

\hypersetup{%
pdfauthor = {\AUTHORS},
pdftitle = {\TITLE},
pdfsubject = {\CONFERENCE},
pdfkeywords = {\KEYWORDS},
bookmarksopen = {true}
}

\definecolor{placeholderbg}{rgb}{0.85,0.85,0.85}

\usepackage[pdftex]{graphicx}
\usepackage{subcaption}
\usepackage{soul}
\soulregister\cite7
\soulregister\ref7
\soulregister\pageref7

\usepackage[lined, ruled]{algorithm}
\usepackage{multirow}
\usepackage{array}	
\usepackage{xspace}
\usepackage{listings}

\definecolor{mygreen}{rgb}{0,0.6,0}
\definecolor{mygray}{rgb}{0.5,0.5,0.5}
\definecolor{mymauve}{rgb}{0.58,0,0.82}

\lstdefinestyle{MyCodeStyle}
{
backgroundcolor=\color{white},   
  basicstyle=\footnotesize,        
  breakatwhitespace=true,         
  breaklines=true,                 
  captionpos=b,                    
  commentstyle=\color{mygreen},    
  escapeinside={!}{!},          
  extendedchars=true,              
  keepspaces=true,                 
  keywordstyle=\color{blue},       
  morekeywords={function,local},            
  numbers=left,
  numberstyle=\footnotesize\color{mygray},
  xleftmargin=1.8em,
  numbersep=0.8em,
  aboveskip=0pt,
  belowskip=0pt,
  rulecolor=\color{black},         
  showspaces=false,                
  showstringspaces=false,          
  showtabs=false,                  
  stringstyle=\color{mymauve},     
  tabsize=2,                       
  title=\lstname,                   
  basicstyle=\ttfamily,
  columns=flexible
}

\usepackage{silence}
\lstdefinelanguage{modp}
{
  morekeywords={
    module,
    machine,
    refines,
    satisfies,
    test,
    hide,
    in,
    sends,
    send,
    creates,
    private,
    receives, 
    implementation,
    specification, 
    event, 
    model, 
    \$,
    bool,
    any, 
    static, 
    fun, 
    if, 
    int, 
    interface,
    entry,
    state, 
    var,
    new,
    goto,
    hot,
    start,
    on,
    monitor,
    observes,
    eventset, 
    is
  },
  sensitive=false, 
  morecomment=[l]{//}, 
  morecomment=[s]{/*}{*/}, 
  morestring=[b]" 
}

\definecolor{eclipseBlue}{RGB}{42,0.0,255}
\definecolor{eclipseGreen}{RGB}{0, 139, 69}
\definecolor{eclipsePurple}{RGB}{127,0,85}

\lstdefinestyle{ModPCodeStyle}
{
  basicstyle=\ttfamily\footnotesize, 
  captionpos=b, 
  extendedchars=true, 
  tabsize=2, 
  columns=fixed, 
  keepspaces=true, 
  showstringspaces=false, 
  breaklines=true, 
  numbers=left,
  xleftmargin=1.8em,
  numbersep=0.8em,
  aboveskip=0pt,
  belowskip=0pt,
  commentstyle=\color{eclipseBlue}, 
  keywordstyle=\bf\color{eclipsePurple}, 
  stringstyle=\color{eclipseBlue}, 
}

\newcommand{\clusterA}{Cluster-A\xspace}
\newcommand{\clusterB}{Cluster-B\xspace}

\newcommand{\imagenet}{ILSVRC12\xspace}

\newcommand{\vgg}{VGG16\xspace}
\newcommand{\inception}{Inception-v3\xspace}
\newcommand{\svt}{S2VT\xspace}

\newcommand{\stage}{stage\xspace}
\newcommand{\stages}{stages\xspace}

\newcommand{\mamfull}{\texttt{NUM\_OPT\_ACTIVE\_MINIBATCHES}\xspace}
\newcommand{\mam}{NOAM\xspace}
\newcommand{\asp}{ASP\xspace}

\titlespacing\section{0pt}{6pt plus 4pt minus 2pt}{3pt plus 2pt minus 2pt}
\titlespacing\subsection{0pt}{6pt plus 2pt minus 2pt}{3pt plus 2pt minus 2pt}
\titlespacing\subsubsection{0pt}{4pt plus 4pt minus 2pt}{2pt plus 2pt minus 2pt}

\usepackage[absolute]{textpos}

\newfloat{acmcr}{b}{acmcr}

\newcommand{\note}[2]{
    \ifthenelse{\equal{\showComments}{yes}}{\textcolor{#1}{#2}}{}
}

\newcommand*\samethanks[1][\value{footnote}]{\footnotemark[#1]}

\usepackage{balance}
\usepackage{stfloats}
\usepackage{hhline}

\newcommand{\cmu}{{\large$^\dag$}}
\newcommand{\stanford}{{\large$^\ddagger$}}
\newcommand{\msr}{{\large$^\star$}}

\date{}
\title{\textbf{\TITLE}}

\author{
Aaron Harlap\cmu\hspace{0.01in}\thanks{Work started as part of an internship at Microsoft Research.} \hspace{0.3in} Deepak Narayanan\stanford\hspace{0.01in}\samethanks \hspace{0.3in}
\\
\vspace{-0.15in}
\\
Amar Phanishayee\msr \hspace{0.2in} Vivek Seshadri\msr \hspace{0.2in} Nikhil Devanur\msr \hspace{0.2in} Greg Ganger\cmu \hspace{0.2in} Phil Gibbons\cmu
\\
\\
\rm{\textit{\msr Microsoft Research \cmu Carnegie Mellon University \stanford Stanford University}}
}

\begin{document}

\maketitle

\ifthenelse{\equal{\PAGENUMBERS}{yes}}{%
  \thispagestyle{fancy}
}{%
  \thispagestyle{empty}
}


\begin{abstract}

\system{} is a Deep Neural Network (DNN) training system for GPUs
that parallelizes computation by pipelining execution across multiple machines.
Its {\it pipeline parallel} computing model avoids the slowdowns faced
by data-parallel training when large models and/or limited
network bandwidth induce high communication-to-computation ratios.
\system{} reduces communication by up to 95\% for
large DNNs relative to data-parallel training, and
allows perfect overlap of communication and computation.
\system{} keeps all available GPUs productive by systematically
partitioning DNN layers among them to balance work and minimize communication,
versions model parameters for backward pass correctness, and
schedules the forward and backward passes of different inputs
in round-robin fashion to optimize ``time to target accuracy''.
Experiments with five different DNNs on two different clusters
show that \system{} is up to 5x faster in time-to-accuracy
compared to data-parallel training.

\end{abstract}

\ifthenelse{\equal{\onlyAbstract}{no}}{%
\section{Introduction}
\label{sec:intro}

The last five years has seen a rapid increase in the use of Deep Neural
Networks (DNNs), with researchers and practitioners applying these models to
great effect across a wide range of applications, including
image and video classification, speech recognition, and language translation~\cite{he2016deep,speechdnn,sent1,conv2,capt1}.
As DNNs have become more widely developed and used, model sizes have grown to increase effectiveness---models today have
tens to hundreds of layers totaling  10--20 million parameters. 
Such growth not only stresses the already time- and resource-intensive DNN
training processes, but also causes the commonly used parallelization
approaches to break down.

The most common approach  is \emph{data parallelism}, 
where the DNN model is replicated on multiple worker machines, 
with each worker processing a subset of the training data.
Weight updates computed on individual workers are aggregated to
obtain a final weight update that reflects updates across all inputs.
The amount of data communicated per aggregation is proportional to the size of the model.
Although data-parallel training works well with some popular models
that have high computation-to-communication ratios, two important trends threaten its efficacy.
First, growing model sizes increase per-aggregation communication.
Indeed, some widely-used models are large enough that the communication overheads already eclipse computation time, 
limiting scaling and dominating total training time (e.g., up to 85\% of training time for VGG-16~\cite{simonyan2014very}).
Second, rapid increases in GPU compute capacity further
shift the bottleneck of training towards communication across models.
Our results show these effects quantitatively (Figure~\ref{fig:communication-ratio})
for three generations of NVIDIA GPUs (Kepler, Pascal, and Volta),
across five different DNN models.

Another approach to distributed training, \emph{model parallelism}, is used traditionally for models that are too large
to keep in a worker's memory or cache during training~\cite{dean2012deepbelief,krizhevsky2014one,chilimbi2014adam}.
Model-parallel training involves partitioning the model among workers such that
each worker evaluates and performs updates for only a subset of the model's parameters.
However, even though model parallelism enables training of very large models,
traditional model parallelism can lead to severe underutilization of compute resources since it 
either actively uses only one worker at a time (if each layer is assigned to a worker) or 
cannot overlap computation and communication (if each layer is partitioned).
In addition, determining how best to partition a DNN model among workers
is a challenging task even for the most experienced machine learning practitioners~\cite{dean2017rlplacement},
often leading to additional inefficiency.

This paper describes \system{}, a new distributed training system specialized for DNNs.
Like model parallelism, it partitions the DNN and assigns
subsets of layers to each worker machine.
But, unlike traditional model parallelism, \system{}
aggressively pipelines minibatch processing, with different workers processing
different inputs at any instant of time.
This is accomplished by injecting multiple inputs
into the worker with the first DNN layer, thereby keeping the pipeline full
and ensuring concurrent processing
on all workers.
It also uses data parallelism for selected subsets of layers to balance computation load among workers.
We refer to this combination of pipelining, model parallelism, and data parallelism
as \emph{pipeline-parallel training}.

Pipeline-parallel training has the potential to provide high DNN training
performance when data parallelism struggles.
In particular, inter-worker communication can be limited to
activations (on the forward pass) and gradients (backward) between
adjacent layers assigned to different workers. We observe such communication to
be up to 95\% less than that for data-parallel training.

\system{} is the first system to combine pipelining, model parallelism, and data
parallelism in a general and automated way.
It realizes the potential of pipeline parallelism with a design that addresses
several challenges.
First, like pipelining in processors, achieving high efficiency
requires the right partitioning of the DNN into ``stages'' (layer sub-sequences)
that are each executed on a different worker;
this depends on both the model architecture and the hardware deployment.
Bad partitionings, where stages have widely skewed amounts of work,
can lead to workers spending significant time idle.
\system{} automatically determines how to partition the layers of the DNN based on a short
profiling run, using an algorithm that balances computation
load among the different stages while minimizing communication.
Second, since DNNs will not always divide evenly among available
workers, \system{} can use data parallelism
for some stages---multiple workers can be assigned to a given
stage, processing different minibatches in parallel.
Third, unlike traditional uni-directional pipelines, DNN training
is bi-directional---the forward pass is followed by a backward pass
through the same layers in \emph{reverse} order.
\system{} interleaves forward and backward minibatch processing on each
worker, while making sure to route minibatches through the same workers on the backward pass.
This helps to keep all workers busy without pipeline stalls,
while preventing excessive in-progress minibatches and ensuring model convergence.
Fourth, weight versions need to be managed carefully to obtain a high-quality model at the end of training.
We find that allowing the backward pass for a given minibatch
to use more up-to-date parameters than those used in the corresponding forward
pass can be a significant problem.
\system{} maintains parameter value versions for each in-flight minibatch to combat this problem.

Experiments with \system{} confirm its effectiveness for each of the five models we evaluated and that cover two important DNN classes 
\--- CNNs and RNNs (seq-to-seq).
When training Inception-v3~\cite{ioffe2015batch}, VGG16~\cite{simonyan2014very}, Resnet-50~\cite{he2016deep}, and AlexNet~\cite{krizhevsky2012imagenet} on the \imagenet{}~\cite{russakovsky2015imagenet} dataset, \system{} speeds up training by up to 1.45x, 5.12x, 1.21x and 6.76x respectively compared to data-parallel BSP. When training the S2VT~\cite{venugopalan2015sequence} model on the MSVD~\cite{chen2011collecting} dataset, \system{} speeds up training by 3x compared to data-parallel BSP.

To summarize, this paper makes four primary contributions.
First, it introduces a parallelization approach specialized to DNN training to address communication bottlenecks
by combining model parallelism with aggressive pipelining and data parallelism where appropriate.
Second, it identifies the key challenges in realizing the performance potential of this idea and details solutions for each.
Third, it describes a system (\system{}) that efficiently implements pipeline-parallel DNN training.
Fourth, it experimentally demonstrates that \system{} allows parallel DNN
training in circumstances where communication overheads limit data-parallel
training, including cases where data-parallel is slower than single-machine training.

\section{Background \& Related Work}
\label{sec:background}

This section discusses distributed DNN training,
including terminology, common approaches and their limitations, and related work,
using training of image classification models as a concrete example.

\begin{figure*}
    \vspace{-0.1in}
    \centerline{\includegraphics[keepaspectratio=1,width=0.95\textwidth]{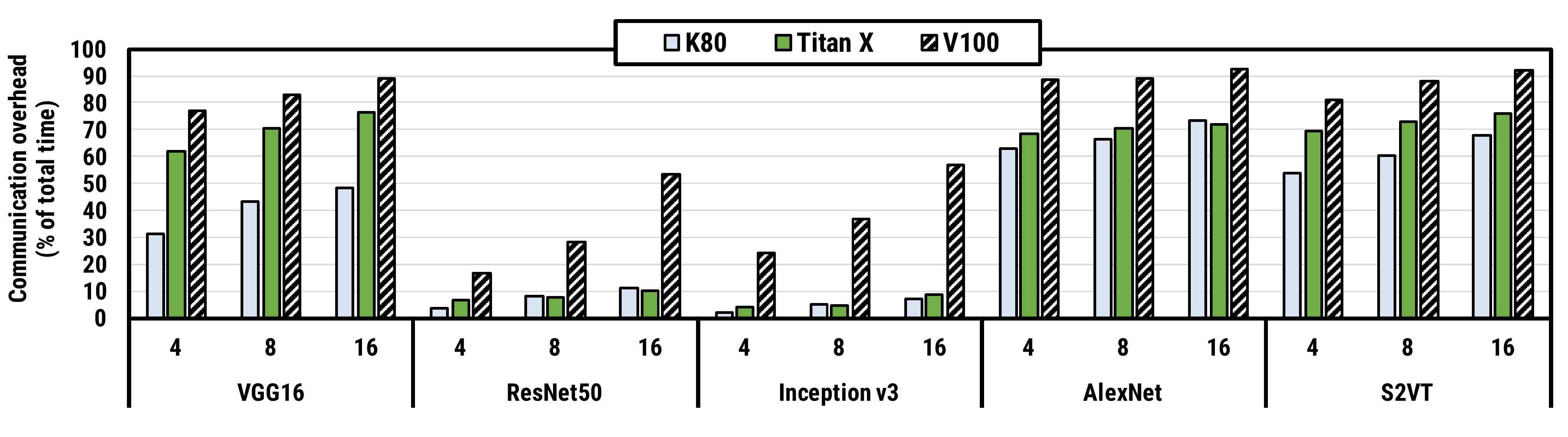}}
    \vspace{-0.1in}
    \caption{
        \small
       Communication overhead as a percentage of total training time for different hardware configurations. Many models (AlexNet, \vgg, \svt) have a high communication overhead, even on the relatively slow K80.  Two factors contribute to an increase in the communication overhead across \textit{all} models: (i) an increase in the number of data-parallel workers, and (ii) an increase in GPU compute capacity.
    } 
    \label{fig:communication-ratio}
    \vspace{-0.1in}
\end{figure*}

\subsection{DNN Training}
\label{sec:training}

A DNN model consists of a sequence of layers of different types (e.g., convolutional, fully connected, pooling). 
DNN models are typically trained using a dataset of labeled images.
Training consists of multiple \emph{epochs}, where an epoch is one iteration
through all images in the dataset. 
In each epoch, the model trains over all images in the dataset in \emph{steps}.
In each step, the current model first makes a prediction for a small set of
training samples, also known as a \emph{minibatch}. This process is referred to
as a \emph{forward pass}. To make a prediction, input data from the minibatch
is fed to the first layer of the model. Each layer then computes a function
over its inputs, often using \emph{learned} parameters (or weights), to produce
an output for the next layer. The output of the last layer is the class
prediction. Based on the model's predicted label and the actual label of each
image, the output layer computes a loss (or error). In the ensuing \emph{backward
pass}, each layer computes 1)~the error for the previous layer, and
2)~the weight update (gradient of the loss) for all relevant layers, which move the
model's predictions toward the desired output.

The goal of DNN training is to obtain a high-accuracy model in as little time as possible. 
This goal can be captured with two metrics: 
1)~\emph{statistical efficiency}, the number of epochs needed to reach a desired level of accuracy,  
and 2)~\emph{hardware efficiency}, the time required to complete a single epoch. 
The total training time to reach a desired accuracy level is simply the product
of these two metrics~\cite{hadjis2016omnivore}.
To train large models in a reasonable amount of time, training is distributed
across multiple GPUs\footnote{For the rest of this paper, we use the terms
``GPU'', ``worker'', and ``machine'' interchangeably, although a machine may
have multiple GPUs each running a worker thread.},
usually using one of two approaches: data or model parallelism.

\paragraph{Data Parallelism.}
With data parallelism, the input dataset is partitioned across multiple GPUs. 
Each GPU maintains a full copy of the model and trains on its own partition of data while periodically synchronizing weights with other GPUs,
using either collective communication primitives~\cite{goyal2017accurate} or Parameter Servers~\cite{li2014parameterserver,cui2016geeps}. 
The frequency of parameter synchronization affects both statistical and hardware efficiency.

On one end, synchronizing at the end of every minibatch (referred to as
\emph{bulk synchronous parallel} or BSP~\cite{valiant1990bmpc}) 
reduces the staleness of weights used to compute gradients, ensuring good statistical efficiency.
However, as shown in Figure~\ref{fig:data_parallel_overview},
BSP requires each GPU to \emph{wait} or \emph{stall} for gradients from other GPUs, 
thus significantly lowering hardware efficiency.  
Despite optimizations such as Wait-free Backpropagation~\cite{PoseidonATC2017},
where weight gradients are sent as soon as they are available,
(common in modern parameter servers),
communication stalls are inevitable in data-parallel training due to the structure
of the DNN computation, and the fact that communication can often dominate
total execution time.
Furthermore, rapid increases in computation speeds further shift the training
bottleneck towards communication.

\begin{figure}[h]
    \centerline{\includegraphics[keepaspectratio=1,width=0.7\columnwidth]{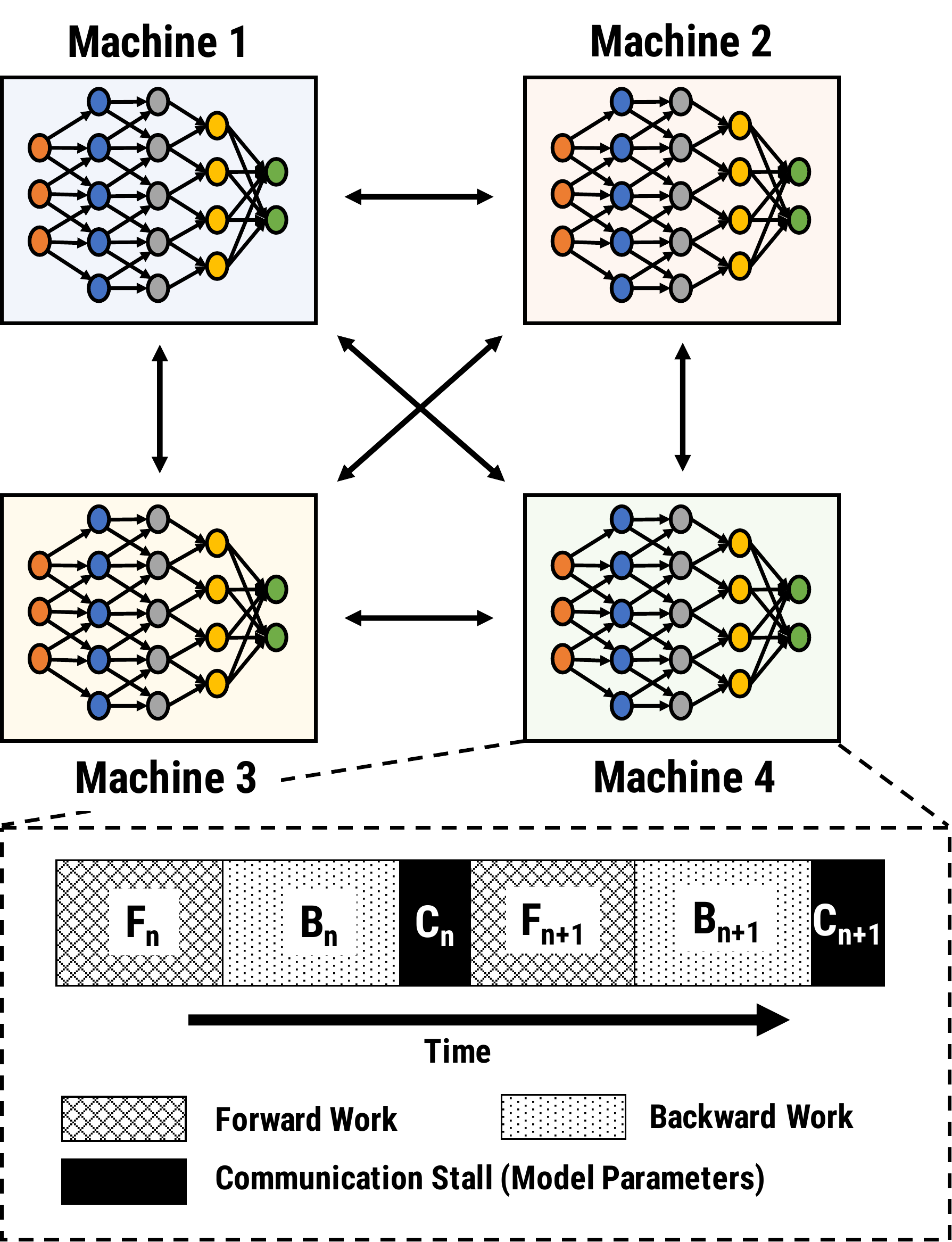}}
    \caption{
        \small
        Example data-parallel setup with 4 machines. Timeline at one of the machines shows communication stalls during model parameter exchanges.
    }
    \label{fig:data_parallel_overview}
    \vspace{-0.1in}
\end{figure}

Figure~\ref{fig:communication-ratio} quantitatively shows the fraction of training time spent in communication stalls for five different DNN models run on ``commodity'' public cloud servers using three different generations of NVIDIA GPUs---Kepler (K80), Pascal (Titan X), and Volta (V100)---linked by a 10Gbps network.
We focus on three takeaways.  First, starting with slower GPUs such as the K80s, we note that some CNNs (\vgg and AlexNet) and the  sequence-to-sequence model for video transcription (\svt) spend significant time communicating.
Networks such as ResNet50 and \inception have comparatively low communication overhead.
Second, as the number of data parallel workers increases, communication overheads increase for all models.
Third, as GPU compute speeds increase (K80s to V100s), communication stalls also increase for all five models.

Prior work has proposed more relaxed synchronization models, 
where each GPU proceeds with the computation for the next minibatch without stalling for the gradients from the other GPUs. 
This approach, which we refer to as \emph{asynchronous parallel} or \asp{}, reduces GPU idle time, 
and as a result, improves hardware efficiency compared to BSP. 
However, this process can lead to gradients being computed on stale weights,
thereby lowering statistical efficiency. 
Our experimental results corroborate recent findings that show that \asp{}
does not reduce end-to-end DNN training time~\cite{cui2016geeps,tensorflow,revisiting-bsp-2016}.

\vspace{-0.1in}
\paragraph{Model Parallelism.}
With model parallelism, the model is partitioned across multiple
GPUs, with each GPU responsible for only a portion of the model.
For machine learning (ML) problems such as matrix factorization, topic modeling, and linear regression, prior work~\cite{low2012distributedgraphlab,lee2014model,yuan2015lightlda,kim2016strads} has shown that model parallelism can often achieve faster training times than data parallelism 
because of the improved statistical efficiency that arises from not
using extremely large minibatch sizes; the STRADS framework~\cite{kim2016strads} shows that
pipelining multiple minibatches can further improve training times for these ML problems.
Model parallelism has also been used for DNNs, but traditionally
only as a last resort when the working set of model training is too large to
fit in a single worker's memory or cache~\cite{krizhevsky2014one,chilimbi2014adam,dean2012deepbelief}
(making data parallelism not an option).
This is because traditional model-parallel DNN training suffers from two major limitations.
  
First, model-parallel DNN training results in severe under-utilization of
GPU resources, as illustrated in Figure~\ref{fig:no_pipeline}.
The figure shows a partitioning of the DNN layers across four machines,
such that each machine is responsible for a group of consecutive layers; in this
regime, the inter-layer values (activations and gradients) between these groups are the only
parameters that need to be communicated across machines.
\footnote{While other partitioning schemes are possible, this is the most
common, and the one we will use for model parallelism in this paper.}
For each minibatch, only a single \stage is active at any instant of time.
Pipelining multiple minibatches back-to-back would improve utilization,
but is traditionally not done because, 1)~the bi-directionality of DNNs
(the forward pass is followed by a backward pass through the same layers in reverse order)
makes pipelining challenging,
and more importantly 2)~a naive pipelining mechanism introduces weight update
computations on \emph{stale} weights, leading to the final model achieving a
lower accuracy than in the data-parallel training.

Second, the burden of partitioning a model across multiple GPUs is left to the
programmer~\cite{krizhevsky2014one}, resulting in point solutions.
Recent work explores the use of reinforcement learning to automatically
determine device placement for model parallelism~\cite{dean2017rlplacement}.
Unfortunately, such online decision making techniques are time- and resource-intensive;
they also don't seamlessly combine pipelining, data-, and model- parallelism.

\begin{figure}[t!]
    \centerline{\includegraphics[keepaspectratio=1,width=0.85\columnwidth]{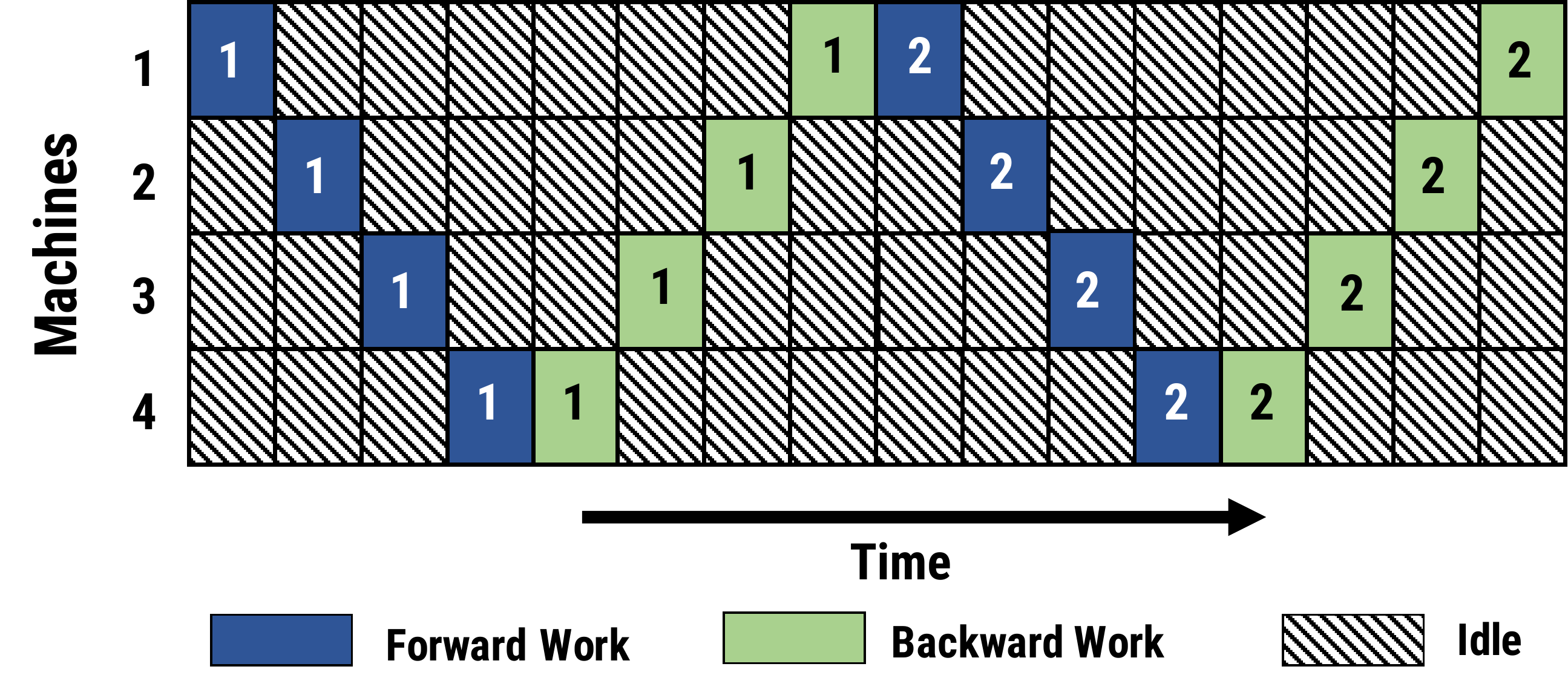}}
    \caption{
       \small
       Model parallel training with 4 machines. Numbers indicate minibatch ID. For simplicity, here we assume that forward and backward work in every stage takes one time unit, and communicating activations across machines has no overhead.
    }
    \label{fig:no_pipeline}
    \vspace{-0.1in}
\end{figure}

\paragraph{Other Related Work.} 
Recent work on fast training of Convolutional Neural Networks (CNNs) make use of highly optimized and expensive clusters with high-speed intra- and inter-machine interconnects~\cite{goyal2017accurate, dgx1}.
Many public cloud providers do not yet offer such optimized server SKUs and one
can expect such offerings to be prohibitively expensive when they are offered.
In contrast, in our work we investigate the use of commodity SKUs available in public cloud offerings;
this is training infrastructure readily accessible to the masses.

CNTK's~\cite{cntk} 1-bit quantization technique tackles the problem of
communication bottlenecks in data parallelism training~\cite{seide20141}.
This approximation strategy lacks generality and is effective for limited scenarios; 
it does not hurt convergence for some speech models~\cite{psgd-speech-2014}, 
but hurts statistical performance due to noisy gradients in many others~\cite{cui2016geeps, tensorflow}.

Goyal et al.~\cite{goyal2017accurate} uses more efficient implementations of
\texttt{all\_reduce}, like the recursive halving-and-doubling algorithm and the bucket algorithm
to reduce the amount of data being sent over the network~\cite{thakur2005optimization}. 
Others have explored techniques from the HPC literature to reduce the overhead
of communication~\cite{baidu_performance:_2017,uber_performance:_2017}. But
all these reduction approaches still involve synchronous communication
patterns, resulting in smaller network stalls that only slightly alleviate the
communication bottleneck introduced due to ever growing model sizes and faster compute capabilities.

Chen et al.~\cite{pbackprop2012} briefly explore the potential benefits of pipelining minibatches in model parallel training, but
do not address the conditions for good statistical efficiency, scale, and generality as applicable to large real-world models.
In fact, our work shows that pipelining computation naively \emph{is not enough} for real-world models.
In our proposed solution (Section~\ref{sec:pipelining}), we address key issues ignored in prior work, and offer a general and automated solution.

\begin{figure}[t!]
    \centerline{\includegraphics[keepaspectratio=1,width=0.7\columnwidth]{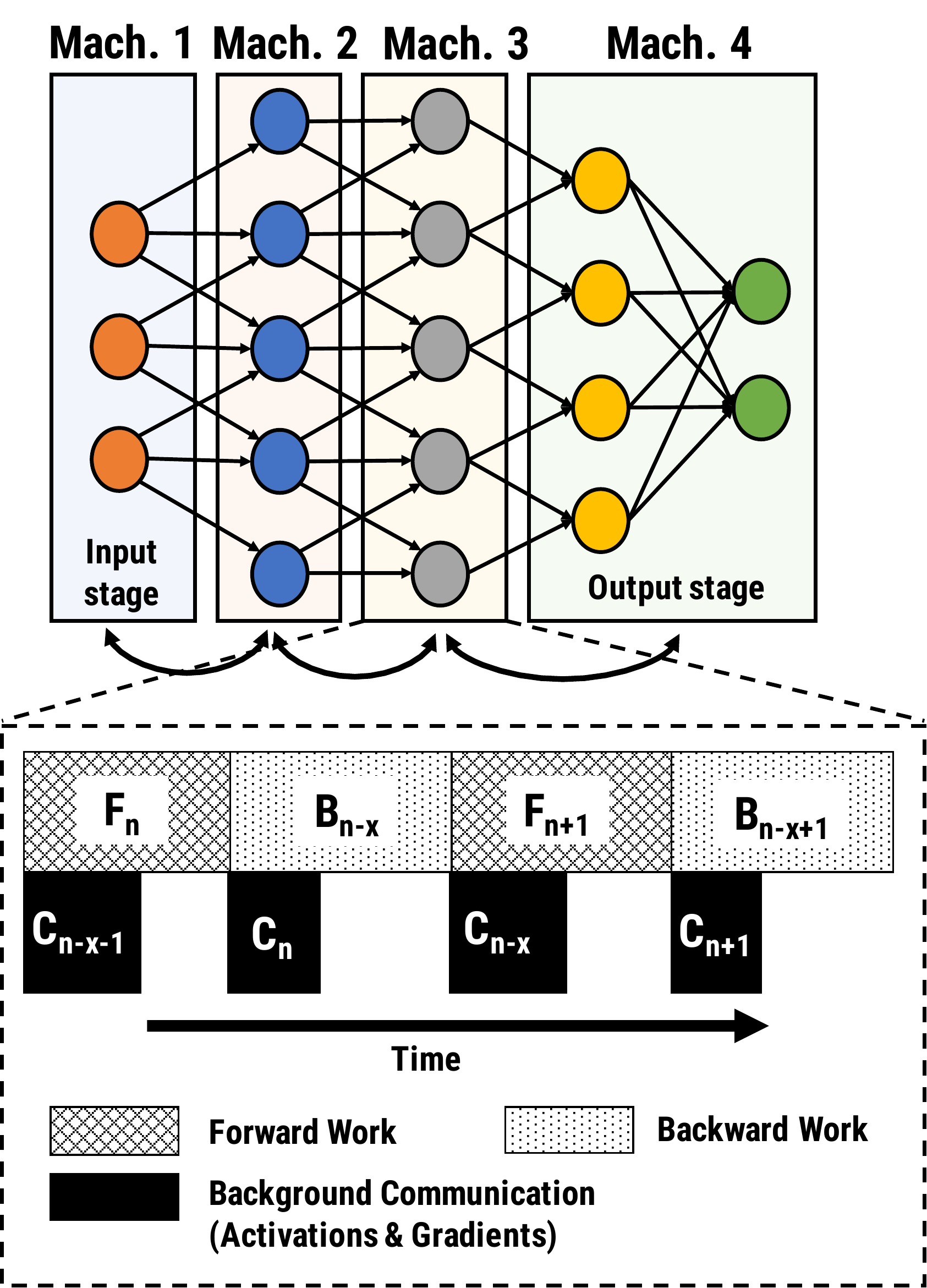}}
    \caption{
        \small
        An example pipeline-parallel assignment with four machines and an example timeline at one of the machines, highlighting the temporal overlap of computation and activation / gradient communication.
    }
    \label{fig:pipeline_parallel_overview}
    \vspace{-0.1in}
\end{figure}

\section{Parallel Training in \system{}}
\label{sec:pipelining}

\system combines traditional data parallelism with model parallelism enhanced with pipelining.  
We call this scheme \emph{pipeline parallelism} (PP). 
In this section, we first describe PP, and then describe \system{}'s design that
addresses the challenges associated with making pipeline-parallel training work effectively.

\subsection{Pipeline Parallelism}
Pipeline-parallel training partitions the layers of the model being trained into
multiple \emph{\stages} -- each \stage contains a \emph{consecutive} set of layers in the model.
Each \stage is mapped to a separate GPU that performs both the forward and
backward pass for all the layers in that \stage. We refer to the \stage that
contains the input layer as the \emph{input \stage}, and the one that
contains the output layer as the \emph{output \stage}.
Figure~\ref{fig:pipeline_parallel_overview} shows a simple example of a
pipeline-parallel assignment, where the DNN is split across four machines.

In the simplest case, only one minibatch is active in the system, as in traditional model-parallel training.
Figure~\ref{fig:no_pipeline} shows the computation timeline of an example configuration with
four machines and one active minibatch in the pipeline.  In the forward
phase, each \stage performs the forward pass for the minibatch for the layers
in that \stage and sends the results to the next \stage. The output \stage,
after completing its forward pass, computes the loss for the minibatch. In the
backward phase, each \stage performs the backward pass and propagates the loss
to the previous \stage. With only one active minibatch, at most one GPU is
active at any given point in time. 

To ensure that no GPU is idle at any point in time, we inject multiple
minibatches into the pipeline one after the other, thus enhancing model-parallel training with pipelining.
On completing the forward pass for a minibatch, each \stage asynchronously sends the output
activations to the next \stage, while simultaneously starting to process
another minibatch. Similarly, after completing the backward pass for a
minibatch, each \stage asynchronously sends the gradient to the previous
\stage, while starting computation for another minibatch. 
Pipelining has two main advantages over data-parallel training:

\noindent\textbf{Pipelining communicates less.} PP requires far less 
communication compared to BSP. Figure~\ref{fig:vgg-layer-size} 
compares the size of each layer's output to the overall size of model parameters for VGG16.
Instead of having to communicate all the parameters, as is done in BSP,
each machine in a PP execution only has to communicate the output data of
\textit{one} of the layers.
This often results in large reductions in communication (e.g., $>$90\% reduction for VGG16). 

\noindent\textbf{Pipelining overlaps computation and communication.} 
Asynchronous communication of forward output activations
and backward gradients across stages
results in a significant overlap of communication with computation of a subsequent minibatch, 
thus achieving better hardware efficiency compared
to BSP (Figure~\ref{fig:pipeline_parallel_overview}).

\begin{figure}[t!]
    \vspace{-0.1in}
    \includegraphics[keepaspectratio=1,width=0.85\columnwidth]{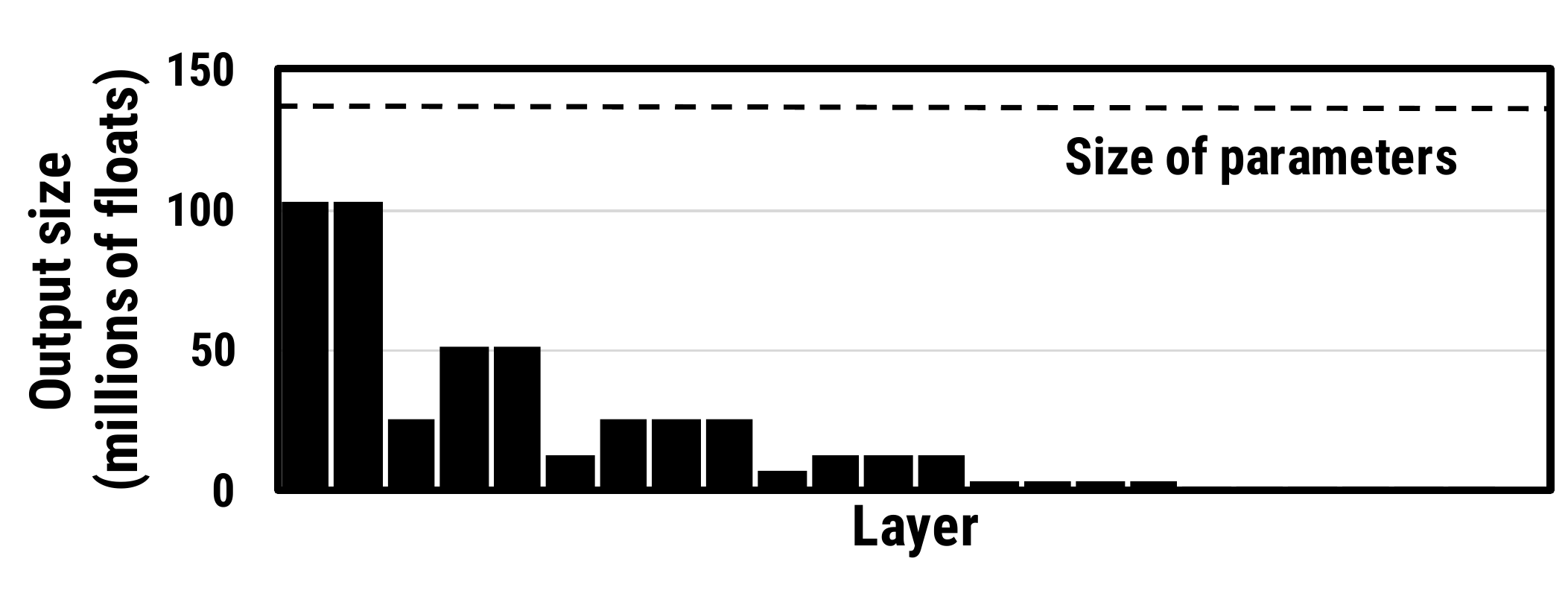}
    \vspace{-0.18in}
    \caption{
        \small
        Sizes of layer output data for VGG16 with a minibatch size of 32 on ImageNet1K data.
        The black dotted line indicates the size of the model parameters.
    }
    \label{fig:vgg-layer-size}
    \vspace{-0.15in}
\end{figure}

While pipelining by itself reduces training time compared to data parallelism, we
observe model parallelism and data parallelism work
best for different types of layers~\cite{krizhevsky2014one}. 
As a result, \system aims to combine pipelined model parallelism and data parallelism in a manner that minimizes overall training time.
Figure~\ref{fig:pipeline_plus_data_parallelism} shows how pipeline parallelism might partition layers of a hypothetical model across stages on 8 machines. 
However, there are three challenges that need to be addressed to make this approach effective for large real-world DNN models:
\begin{enumerate}
\item Automatic partitioning of work across available compute resources.
\item Scheduling of computation to maximize throughput while ensuring forward progress in the learning task.
\item Ensuring that learning is effective in the face of asynchrony introduced by pipelining.
\end{enumerate}
The remainder of this section describes these challenges and \system{}'s approach to address them.

\begin{figure}[h]
    \centering
    \includegraphics[width=0.85\columnwidth]{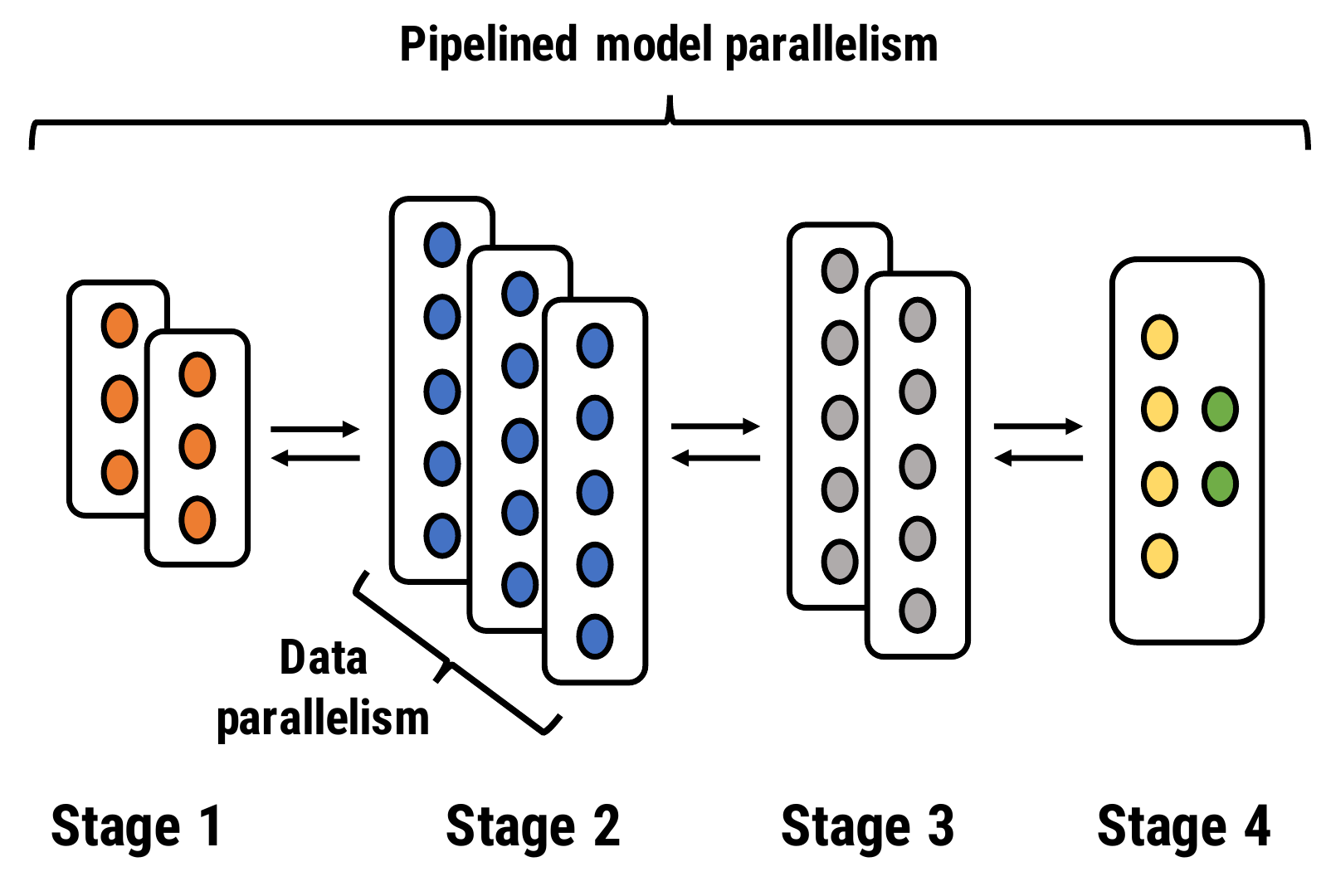}
    \caption{
        \small
        Pipeline Parallel training in \system{} combines pipelining, model- 
        and data-parallel training.
    }
    \label{fig:pipeline_plus_data_parallelism}
    \vspace{-0.1in}
\end{figure}


\subsection{Partitioning Layers Across Machines}
\label{sec:layer-partitioning}

Given a model and a set of machines, \system{}'s first challenge is to automatically
partition layers of the model across available machines so as to minimize overall training time.
Figure~\ref{fig:layer-partitioning} shows the workflow adopted by \system to
partition the layers of the DNN among the available machines.
When partitioning layers into different stages across machines, \system's partitioning algorithm must ensure
that each \stage roughly performs the same amount of total work.
At the same time, the partitioning algorithm must also ensure that the amount of data 
communicated across \stages is as small as possible, to avoid communication stalls. 
Load imbalance across machines or excessive communication between machines can lower hardware efficiency (throughput).

\begin{figure}[t]
    \centering
    \includegraphics[width=1.0\columnwidth]{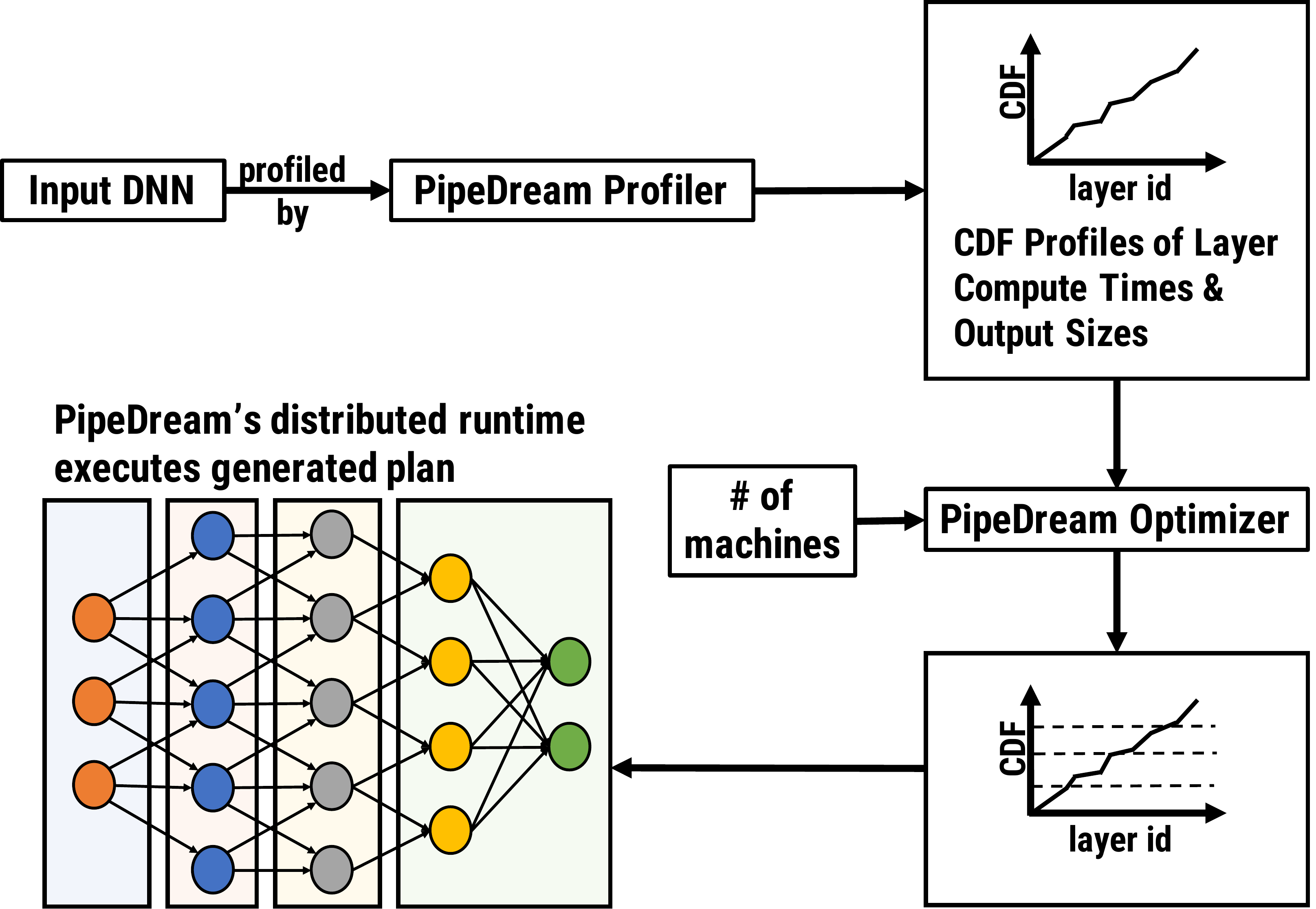}
    \caption{
        \small
        \system's automated mechanism to partition DNN layers into \stages. \system{} first profiles the input DNN, to get estimates for each layer's compute time and output size. Using these estimates, \system{}'s optimizer partitions layers across  available machines.}
    \label{fig:layer-partitioning}
    \vspace{-0.1in}
\end{figure}

Taking these factors into account, given a DNN with $N$ layers and $M$ available
machines, \system first profiles the model on a single machine, and then runs
a partitioning algorithm that groups layers into stages,
while also determining the replication factor for each stage that minimizes the
overall training time for the model.

\textbf{Profiling the DNN Model.} 
Our profiling mechanism exploits the fact that DNN training shows little
variance in the computation and communication time across minibatches. \system
records three quantities for each layer $l$: 1)~$T_l$, the total computation
time across the forward and backward pass for the layer, 2)~$a_l$, the size
of the output activations of the layer (also the size of input gradients in the
backward pass), and 3)~$w_l$, the size of parameters for layer $l$.

To determine $T_l$ for all layers, \system profiles a short run of the DNN model
using 1000 minibatches on one of the machines.\footnote{All the GPUs used in
individual experiments are identical. As a result, it is sufficient to profile
performance on a single GPU.} Using this profile, \system computes $T_l$ as
the sum of the forward and backward computation times for the layer $l$.

All communication happens in three steps: 1)~move data from the GPU to the CPU
of the sender, 2)~send data from sender to receiver over the network, and 3)~move
data from the CPU to the GPU of the receiver. \system estimates the time taken
for communication as the amount of data that needs to be transferred divided by
the network bandwidth on the communication link. $C_l$, the time taken to
communicate the activations from layer $l$ to $l+1$ in a pipeline, is estimated
using $a_l$. The amount of data communicated per worker in data-parallel configurations
with $m$ machines is $4\times(m-1)\times|w_l|/m$; this is used to estimate $W_l^m$, the time for weight synchronization for the layer when using a distributed parameter server.

\textbf{\system{}'s Partitioning Algorithm.}
\label{sec:partition-algorithm}
Our partitioning algorithm takes the output of the profiling step, 
and computes: 1)~a partitioning of layers into stages, 2)~the replication factor for
each stage, and 3)~optimal number of minibatches to keep the training pipeline busy. 

The partitioning algorithm tries to minimize the overall training time of the model.
For a pipelined system, this problem is equivalent to minimizing the time taken by the slowest stage of the pipeline.
This problem has the optimal sub-problem property; a pipeline that maximizes throughput given
a machine count is composed of sub-pipelines that maximize throughput for smaller
machine counts. Consequently, we can find the optimal solution to this problem using Dynamic Programming.

Let $A(j,m)$ denote the time taken by the slowest stage in the optimal pipeline between layers $1$ and $j$ using $m$ machines. 
The goal of our algorithm is to find $A(N,M)$, and the corresponding partitioning.
Let $T(i \rightarrow j, m)$ denote the time taken by a single stage spanning layers $i$ through $j$, replicated over $m$ machines.
\begin{equation*}
    T(i \rightarrow j, m) = \frac{1}{m} \max \bigg( \sum_{l=i}^j T_l , \sum_{l=i}^j W_l^m \bigg)
\end{equation*}
where the left term inside the $\max$ is the total computation time for all the
layers in the stage, and the right term is the total communication time for all the layers in the stage.

The optimal pipeline consisting of layers from $1$ through $j$ using $m$ machines
could either be a single stage replicated $m$ times, or be composed of multiple stages.

\noindent\textbf{Case 1:} The optimal pipeline contains only one stage, replicated $m$ times. In this case,
\begin{equation*}
    A(j, m) = T(1 \rightarrow j, m)
\end{equation*}

\noindent\textbf{Case 2:} The optimal pipeline contains more than one stage. In this case, it can be broken into an optimal sub-pipeline consisting of layers from $1$ through $i$ with $m-m'$ machines followed by a single stage with layers $i+1$ through $j$ replicated over $m'$ machines. Then, using the optimal sub-problem property, we have 
\begin{equation*}
    A(j,m) = \min_{1 \leq i < j} \min_{1 \leq m' < m} \max 
    \begin{cases}
    A(i,m-m')\\
    2 \cdot C_i\\
    T(i+1 \rightarrow j, m')
    \end{cases}
    \vspace{-0.05in}
\end{equation*}
where the first term inside the max is the time taken by the slowest stage of the optimal sub-pipeline between layers 1 and $i$ with $m-m'$ machines, the second term is the time taken to communicate the activations and gradients between layers $i$ and $i+1$, and the third term is the time taken by the single stage containing the remaining layers in a data-parallel configuration of $m'$ machines.

\noindent\textbf{Initialization.} $A(1, m) := T(1 \rightarrow 1, m)$, where $T(.)$ is as
defined above, and $m$ is varied from $1$ through $M$ (the total number of machines).
$A(i, 1) := T(1 \rightarrow i, 1)$, where $i$ is varied from $1$ through $N$
(the total number of layers in the model).

\noindent\textbf{Runtime Analysis.} The total number of sub-problems is $O(NM)$.
Time complexity per sub-problem is also $O(NM)$, leading to a total time
complexity of $O(N^2M^2)$.

Based on the partitioning generated by our algorithm, the optimal number of minibatches admitted per input stage to keep the pipeline full in steady state is given by
\centerline{$\lceil$ (\# machines) / (\# machines in the input stage)  $\rceil$.} 
We refer to this quantity as the \mamfull (\mam).

\subsection{Work Scheduling}
Unlike traditional uni-directional pipelines, pipelined DNN training involves a bi-directional pipeline. The forward pass for a minibatch starts at the input layer and the backward pass ends at the input layer. Consequently, each active minibatch in the pipeline may be in a different layer, either in the forward pass or backward pass. As a result, each machine in the system has to make a choice between two options: i) perform the forward pass for a minibatch, thus pushing the minibatch to downstream machines, and ii) perform the backward pass for a different minibatch, thus ensuring forward progress in learning.

A simple scheduling mechanism that always prioritizes forward work hinders overall forward
progress as weight updates can be applied only once backward passes complete. Similarly,
always prioritizing backward work may periodically result in idle machines with no available work.
We propose a scheduling mechanism that avoids these problems.

In the startup phase, the input \stage admits \mam minibatches to keep the pipeline full in steady state.
Once in steady state, each \stage \emph{alternates} between performing the forward and backward pass for a minibatch. 
We call this mechanism \emph{one-forward-one-backward} (1F1B). 
In a balanced pipeline, 1F1B ensures that no GPU is idle in steady state and
that we make forward progress in learning from each minibatch.

Figure~\ref{fig:example_pipeline} shows the corresponding compute timeline for a pipeline with 4 stages each running on one machine. The \mam for this configuration is 4.
In the startup phase, the input \stage admits exactly four minibatches that propagate their way to the output \stage. 
As soon as the output stage completes the forward pass for the first minibatch, it performs the
backward pass for the same minibatch, and then starts alternating between
performing forward and backward passes for subsequent minibatches. 
As the backward pass starts propagating to earlier stages in the pipeline, 
every stage starts alternating between forward and backward pass for different minibatches. 
As shown in the figure, in the steady state, every machine is busy
either doing the forward pass or backward pass for a minibatch. 
For 1F1B to be effective, it is \emph{not necessary} for the forward pass to
take as long as the backward pass.
In fact, we observe that in practice, the backward pass is always larger than the forward pass,
and 1F1B remains an effective scheduling mechanism. 

\begin{figure}[t!]
    \centerline{\includegraphics[keepaspectratio=1,width=0.85\columnwidth]{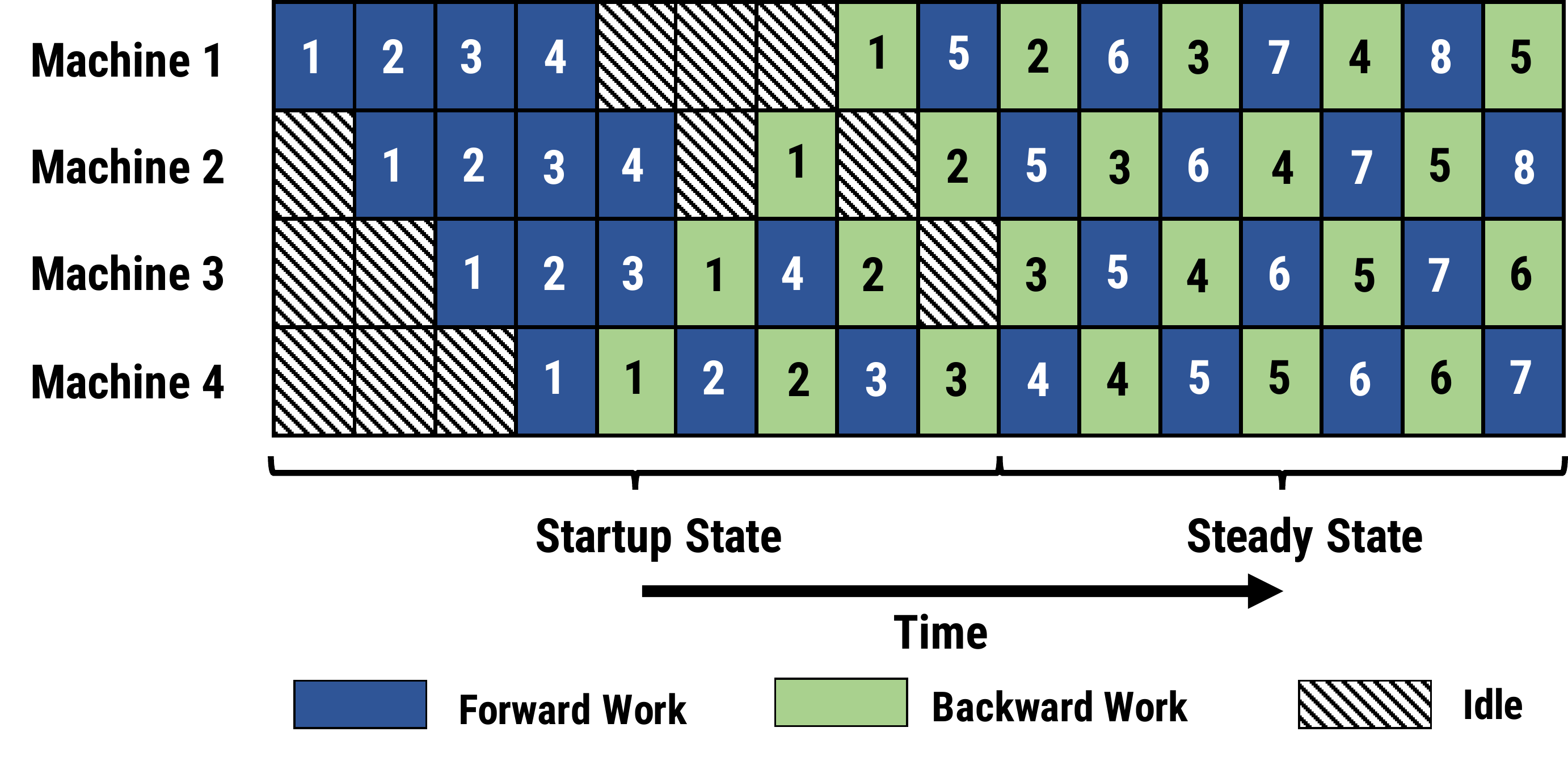}}
    \caption{
        \small
        An example pipeline with 4 machines, showing startup and steady states.
    }
    \label{fig:example_pipeline}
    \vspace{-0.1in}
\end{figure}

When \stage{}s run in a data-parallel configuration, replicated across multiple GPUs, 
we use deterministic round-robin load balancing (\texttt{minibatchID $\mod$ stageReplicaID}) 
to spread work from the previous stages across the replicas.
Such deterministic load-balancing ensures that the backward pass for a minibatch
is performed on the machine responsible for the minibatch's forward pass.

Both the 1F1B scheduling policy for stages in a pipeline and the round-robin scheduling policy for load balancing across replicated stages are \emph{static} policies. 
Thus, they can be executed by each machine independently without requiring expensive distributed coordination.

\subsection{Effective Learning}
\label{sec:avoiding-staleness}

In a naively pipelined system, the forward pass for each minibatch is performed using one version of parameters and the backward pass using a different version of parameters. Figure~\ref{fig:example_pipeline} illustrates this using a partitioning with no data parallelism. If we observe \stage 1 (machine 1), the forward pass for minibatch 5 is performed after the updates from minibatch 1 are applied, whereas the backward pass for minibatch 5 is performed after updates from minibatches 2, 3, and 4 are applied. As a result, in the backward pass for minibatch 5 on \stage 1, the gradient is computed using a different set of weights than the ones used in the corresponding forward pass;  this discrepancy in weight versions can prevent the model from converging. 

Furthermore, different \stages in the DNN model suffer from different degrees of staleness. For example, in the third \stage, each minibatch has only one interleaved update between its forward and backward pass, while the output \stage has no interleaved updates. This asymmetry across layers can further impact model convergence.
Our experimental results show that naive pipelining \emph{does not} achieve
the same accuracy as data-parallel training.
To address this problem, \system uses two techniques. 

\noindent\textbf{Weight Stashing.} Weight Stashing maintains multiple versions of the 
weights, one for each active minibatch. When performing the forward pass, each
\stage processes a minibatch using the latest version of weights available.
After completing the forward pass, \system stores the weights used as part
of the \emph{intermediate} state for that minibatch. When performing the minibatch's
backward pass, the \emph{same version} of the weights is used to compute the
weight gradient.

Weight stashing ensures that \emph{within a stage}, the same version of model parameters
are used for the forward and backward pass of a given minibatch.
For example, in Figure~\ref{fig:example_pipeline}, 
minibatch 5 uses parameters updates from batch 1 on machine 1 and from 2 on machine 2.
Weight stashing says nothing about the consistency of parameter versions used for a given minibatch \textit{across}
\stages.

\noindent\textbf{Vertical Sync.} 
Vertical Sync eliminates the potential inconsistency \emph{across \stages}.  
For example, in Figure~\ref{fig:example_pipeline}, using vertical sync, minibatch 5
uses parameters updated by minibatch 1 on all machines for both its forward and backward passes.
Each minibatch ($m_{i}$) that enters the pipeline
is associated with the latest weight version ($w^{(i-x)}$) seen at the input
\stage. This information is propagated along with the activations and
gradients as the minibatch $m_{i}$ flows through the pipeline in the forward
direction. Across all stages, the forward pass for $m_{i}$ uses the stashed
weights $w^{(i-x)}$, as opposed to the latest weight update.  After performing
the backward pass for $m_{i}$ (using stashed weights $w^{(i-x)}$), each \stage
independently applies weight updates to create the latest weights ($w^{(i)}$),
and can then delete $w^{(i-x)}$. This coordination across stages is asynchronous.

\textbf{Staleness.}
We can now formalize the degree of staleness of
weight updates for each of these techniques. For this discussion,
we assume a straight pipeline with the model split into $n$ stages; the weights
in each stage are represented as $w_1$, $w_2$, and so on. In addition,
we denote $w_1^{(t)}$ as the weights $w_1$ after $t$ minibatches.

Now, after every minibatch, we compute the gradient $\triangledown f(w_1, w_2, \ldots, w_n)$
averaged over all samples in the minibatch.
Vanilla minibatch SGD ($f$ is the loss function we're trying to optimize
and $\nu$ is the learning rate) has the following gradient update,

\vspace{-0.12in}
$$w^{(t+1)} = w^{(t)} - \nu \cdot \triangledown f(w_1^{(t)}, w_2^{(t)}, \ldots, w_n^{(t)})$$
\vspace{-0.2in}

With weight stashing,
gradients in stage $1$ are computed with weights that
are $n$ steps delayed, gradients for stage $2$ are computed with weights that
are $n-1$ steps delayed, and so on. Mathematically, this means our weight update
looks like,

\vspace{-0.15in}
$$w^{(t+1)} = w^{(t)} - \nu \cdot \triangledown f(w_1^{(t-n+1)}, w_2^{(t-n+2)}, \ldots, w_n^{(t)})$$
\vspace{-0.2in}

Without weight stashing, the weight update is not a valid
gradient of the loss function $f$ for any weight vector $w_1, w_2, \ldots, w_n$.

Adding vertical sync alters the weight update to,

\vspace{-0.15in}
$$w^{(t+1)} = w^{(t)} - \nu \cdot \triangledown f(w_1^{(t-n+1)}, w_2^{(t-n+1)}, \ldots, w_n^{(t-n+1)})$$
\vspace{-0.2in}

This is semantically the same as data
parallelism with BSP synchronization on $n$ machines (with the same
original minibatch size on each machine).

\emph{Weight stashing} is critical for meaningful learning.\footnote{In our experiments, we find that the impact of \emph{vertical sync} is negligible.
\system{}'s default semantics exclude vertical sync as it requires more metadata to be stored at every \stage in the pipeline.}
\system{}'s default semantics (weight stashing but
no vertical sync) are between regular minibatched SGD on a single
machine, and data parallelism with BSP synchronization~\cite{cui2014exploiting, ho2013more}.
Our evaluation demonstrates its effectiveness across several models, datasets, and hardware configurations.

\subsection{GPU Memory Management}
\label{sec:memory-management}
As minibatches enter and leave the pipeline, the system has to ensure that the inputs, weights, 
and other intermediate state required by the GPU for its computation are present in GPU memory.
If not managed carefully, the overhead of dynamic memory allocation in the GPU,
and data transfer between GPU and CPU memory can greatly reduce hardware efficiency.

\system extracts the layer parameters from the DNN model and
computes the size of activations, parameters, and intermediate state that needs
to be stored at each \stage, across the active minibatches present in the
pipeline. The number of minibatches for which each \stage has to
maintain intermediate state varies from \stage to \stage. While the output
\stage has to maintain intermediate state for only one active minibatch,
the input \stage needs to do so for \mam minibatches.
\system allocates all required GPU memory at the
beginning of training, and reuses the allocated memory as appropriate.
This significantly reduces the overhead of dynamically managing GPU memory.

\section{Implementation}
\label{sec:implementation}


Figure~\ref{fig:layer-partitioning}
shows \system's high-level workflow.  The input to our system is a model architecture,
the training dataset, and the number of GPUs that will be used for training.
\system first profiles the model on a single machine with a subset of
minibatches from the training dataset. It then runs the optimization algorithm
described in Section~\ref{sec:layer-partitioning} to partition the DNN model into
$k$ \stages, with some stages replicated. The \system runtime then assigns each \stage to a single GPU.

Figure~\ref{fig:implementation} shows the high-level architecture of the \stage
runtime in \system. The interface to \system is implemented as a C++ library
that manages the parameter and intermediate data for the ML worker that
runs on the GPU. In our current implementation, we use Caffe~\cite{jia2014caffe}
as the ML worker. However, \system is extensible and can work with other ML frameworks
such as Tensorflow~\cite{tensorflow}, MXNet~\cite{mxnet}, and CNTK~\cite{cntk} as well.

\begin{figure}
    \centering
    \includegraphics[width=0.8\columnwidth]{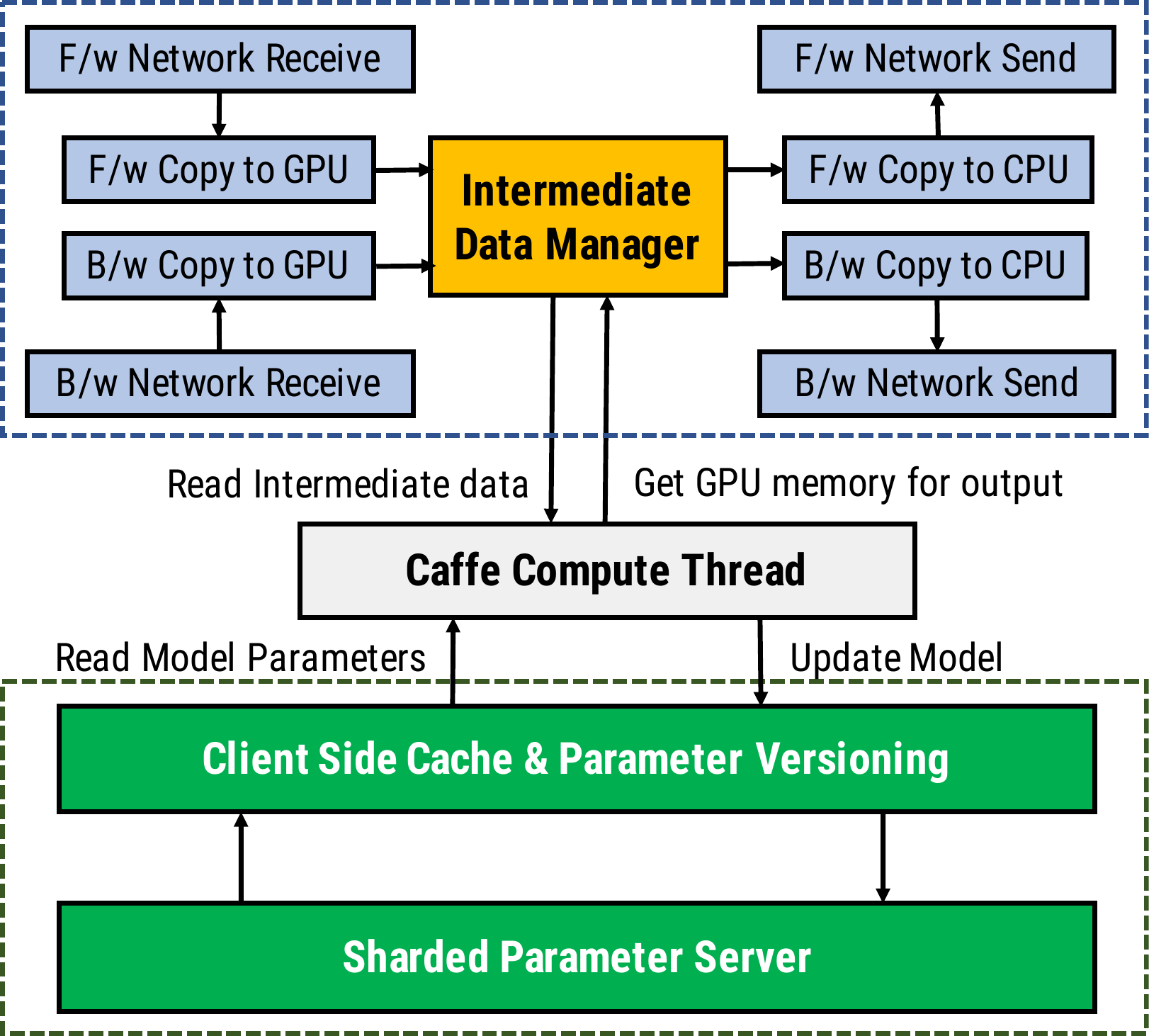}
    \caption{
        \small
        Architecture of the \stage runtime and integration with Caffe. \system{} provides the ML worker thread (Caffe) pointers to GPU memory containing layer input data, parameters, and buffers for recording layer outputs and parameter updates. \system{} manages buffer pools, and handles all intra- and inter-machine communication.}
    \label{fig:implementation}
    \vspace{-0.15in}
\end{figure}

As an initialization step, the \system library in each machine initializes the
GPU data structures corresponding to the \stage that is assigned to the machine.
This involves 1)~initializing the ML worker with the layers that it must
execute as part of the \stage, and 2)~statically allocating memory in the
GPU for the activations, weights, gradients, and the
intermediate state (which includes the input activations and stashed weights
for each active minibatch). Once a machine is initialized, the ML worker pulls its next work assignment
from \system; \system's runtime provides the ML worker with pointers to input data.
The input \stage kick starts the pipeline by creating a minibatch of forward work for its ML worker. 
From then on, each machine follows the 1F1B scheduling algorithm,
while limiting the total number of active minibatches to \mam.

For the assigned minibatch (forward or backward), the ML worker iterates through each layer in the \stage and performs the relevant work for the minibatch.
The ML worker uses appropriate \system API calls to get pointers to the inputs, parameters, outputs, gradients, and intermediate state for each layer. 
Once the minibatch is processed, the ML worker indicates the completion of work to \system, and pulls its next work item.

\textbf{Parameter State.}
For each \stage{}, \system{} maintains all parameters associated with the
layers assigned to the \stage{} directly in GPU memory. The parameters for each
layer are stored separately, and each assigned a unique ID. If the \stage{} is
not replicated, \system{} applies the updates to the most recent version of the
parameter data stored in GPU memory when the weight update is available in the provided
GPU buffer. If the \stage{} is
replicated, the weight update is copied to host memory and then sent to the parameter server.
When a newer version of the parameters becomes available, the prior version is
\emph{not} immediately discarded, as part of the weight stashing scheme.
Parameter data is only discarded once a backward pass that uses fresher
parameters is performed.

\textbf{Intermediate State.} 
Each layer's intermediate data is also assigned a unique blob ID.  Upon receiving
intermediate data from the prior \stage{} (or from disk in the case
of the input stage), \system{} copies the intermediate data to GPU memory and
places a pointer to the associated buffer in a work queue.
Intermediate data from the forward pass is not discarded until the associated
minibatch completes that \stage{}'s backward pass. Intermediate data from
the backward pass is released as soon as the ML worker finishes using it, and
if necessary, after it is sent to the next \stage{}. Due to the differing
requirements for intermediate data in the forward and backward pass, \stages{} in
\system{} commonly manage multiple versions of intermediate data from forward
passes, and just a single version of intermediate data from the currently
running backward pass.

\textbf{Data Parallelism.}
\system uses a distributed parameter server, similar to GeePS~\cite{cui2016geeps},
to synchronize parameters for layers of data-parallel stages.
Using wait-free back propagation, weight gradients are communicated to servers
as soon as they are as computed, rather than waiting for computation to finish
for all layers. 
Each worker contains an instance of a parameter server shard that stores a unique subset
of the parameters. 
The server shards push the newest version of their parameters to the other shards
as soon as updates from all \stage{} replicas are aggregated.

Since we support replication of individual \stages, data-parallel training can
be thought of as a special case in our framework -- we represent this as a
single \stage that contains all the layers of the DNN model, and replicate the
\stage across all available machines.

All inter-machine communication between \system{}'s \stages{}, both in data-parallel
and pipeline-parallel settings, uses ZeroMQ~\cite{zmq} and an efficient
communication stack with fast custom serialization. 

\textbf{Checkpointing.} \system{} supports periodic checkpointing of model
parameters for fault-tolerance, with default checkpointing across stages at the
end of every epoch.  Checkpoints don't require expensive global coordination;
each \stage locally decides to dump its model parameters when it performs the
backward pass for the last minibatch in an epoch. Restarting a failed training
run due to a stage failure entails starting from the last epoch successfully
checkpointed by all the stages.

\section{Evaluation}
\label{sec:eval}

This section compares the effectiveness of \system with data-parallel training
and model-parallel training across models on two clusters. The results of our
experiments support a number of important findings: 1)~combining pipelining,
model parallelism, and data parallelism performs
significantly better than using model parallelism or data parallelism alone,
2)~\system greatly reduces the overhead of communication compared to data-parallel
training, and 3)~\system{}'s improvements are higher for configurations that
have a lower computation-to-communication ratio.

\subsection{Experimental Setup}
\textbf{Datasets.}
We used two datasets in our experiments. The first is the dataset for the Large
Scale Visual Recognition Challenge 2012 (\imagenet{})~\cite{russakovsky2015imagenet}, 
also called the ImageNet 1K dataset.
This dataset has $\sim$1.3 million training images categorized into 1000 classes, and 50,000 validation images. 
The second is the Microsoft Video description corpus (MSVD)~\cite{chen2011collecting}, 
which is a collection of YouTube clips depicting different activities, collected on Mechanical Turk.
The dataset contains 1,970 videos and a vocabulary of 12,594 words to describe them.

\textbf{Clusters.}
We used two different clusters in our experiments. \textit{\clusterA{}} is a private cluster of NVIDIA Titan X GPUs with 12~GB of GPU device memory. Each machine has a E5-2698Bv3 Xeon CPU with 64~GB of RAM. The machines are connected via a 25~Gbps Ethernet interface. \textit{\clusterB{}} is public cloud cluster (AWS p3.2xlarge instances) of NVIDIA V100 GPUs, with 16~GB of GPU device memory. Each machine has a E5-2690 Xeon CPU, 64~GB of RAM with a 10~Gbps Ethernet interface.
Machines on both clusters run 64-bit Ubuntu 16.04 with CUDA toolkit 8.0 and cuDNN v6.

In comparison to \clusterA, \clusterB has faster GPUs, but a slower network.
As a result, as we show later in this section, the performance benefits of
\system are higher in \clusterB than in \clusterA.

\textbf{Models and Training Methodology}
We used three different DNN models in our experiments: 1)~VGG16~\cite{simonyan2014very} with a model size of 550~MB, 2)~Inception-v3~\cite{ioffe2015batch} with a model size of 157~MB, and 3)~S2VT~\cite{venugopalan2015sequence}, a sequence-to-sequence model for video transcription, with a model size of 349MB. We use the \imagenet dataset to train VGG16 and Inception-v3, and the MSVD dataset to train S2VT model. In our experiments, we trained the VGG16 and S2VT models using SGD with a momentum of 0.9 and an initial learning rate of 0.01. For Inception-v3, we used RMSProp~\cite{tieleman2012lecture} with an initial learning rate of 0.045, decayed every two epochs using an exponential rate of 0.94. We used a mini-batch size of 32 per machine for VGG16 and Inception-v3 and a mini-batch size of 80 per machine for S2VT. For all the experiments, we measure the time taken to train the models until they reach their \textit{advertised validation accuracy}: top-1 accuracy of 68\% for VGG16, top-1 accuracy of 67\% for Inception-v3, and METEOR~\cite{denkowski2014meteor} score of 0.294 for S2VT. Guided by prior work, we adjust the learning rate during training to converge to the desired result faster~\cite{movckus1975bayesian, snoek2012practical, sparks2015automating, duchi2011adaptive, kingma2014adam}.

\textbf{\system{}'s Data-Parallel Implementation.}
To measure the performance improvements introduced from using 
\system{}, we compare to \system{}
in data-parallel and single machine configurations.
This allows us to perform a fair comparison of the different schemes.
To confirm its efficiency, we compared \system{}'s
data-parallel implementation to the open source 
version of GeePS, an efficient data-parallel DNN training system that 
also uses Caffe at individual workers and performs favorably, if not
better than, other state-of-the-art DNN frameworks~\cite{cui2016geeps}. 
In our experiments, \system{}'s data-parallel configuration ran at least as
fast as GeePS for all models and datasets tested.

\begin{table*}[t]
\centering
\footnotesize
\begin{tabular}{lrrrrrr}
\toprule
DNN     &  \# Machines & BSP speedup    & \system & \system{} speedup  & \system{} speedup   & \system{} communication  \\
Model   &  (Cluster)   & over 1 machine & Config & over 1 machine     & over BSP  & reduction over BSP\\
\toprule
\multirow{4}{*}{\vgg}  & 4 (A)  & $1.47\times$ & 2-1-1   & $3.14\times$ & $2.13\times$ & 90\%\\
                       & 8 (A)  & $2.35\times$ & 7-1     & $7.04\times$ & $2.99\times$ & 95\%\\
                       & 16 (A) & $3.28\times$ & 9-5-1-1 & $9.86\times$ & $3.00\times$ & 91\%\\\
                       & 8 (B)  & $1.36\times$ & 7-1     & $6.98\times$ & $5.12\times$ & 95\%\\
\midrule
\multirow{2}{*}{\inception}  & 8 (A) & $7.66\times$ & 8 & $7.66\times$ & $1.00\times$ & 0\%\\
                             & 8 (B) & $4.74\times$ & 7-1 & $6.88\times$ & $1.45\times$ & 47\%\\
\midrule
\svt  & 4 (A) & $1.10\times$ & 2-1-1 & $3.34\times$ & $3.01\times$ & 95\%\\
\bottomrule
\end{tabular}
\caption{\small Summary of results comparing \system with data-parallel configurations (BSP) when training models to their advertised final accuracy. ``\system config'' represents the configuration generated by our partinioning algorithm---e.g., ``2-1-1'' is a configuration in which the model is split into three stages with the first stage  replicated across 2 machines.}
\label{table:eval-summary}
\end{table*}

\begin{figure*}[t]
    \centering
    \begin{subfigure}{0.45\textwidth}
    \includegraphics[keepaspectratio=1,width=1\columnwidth]{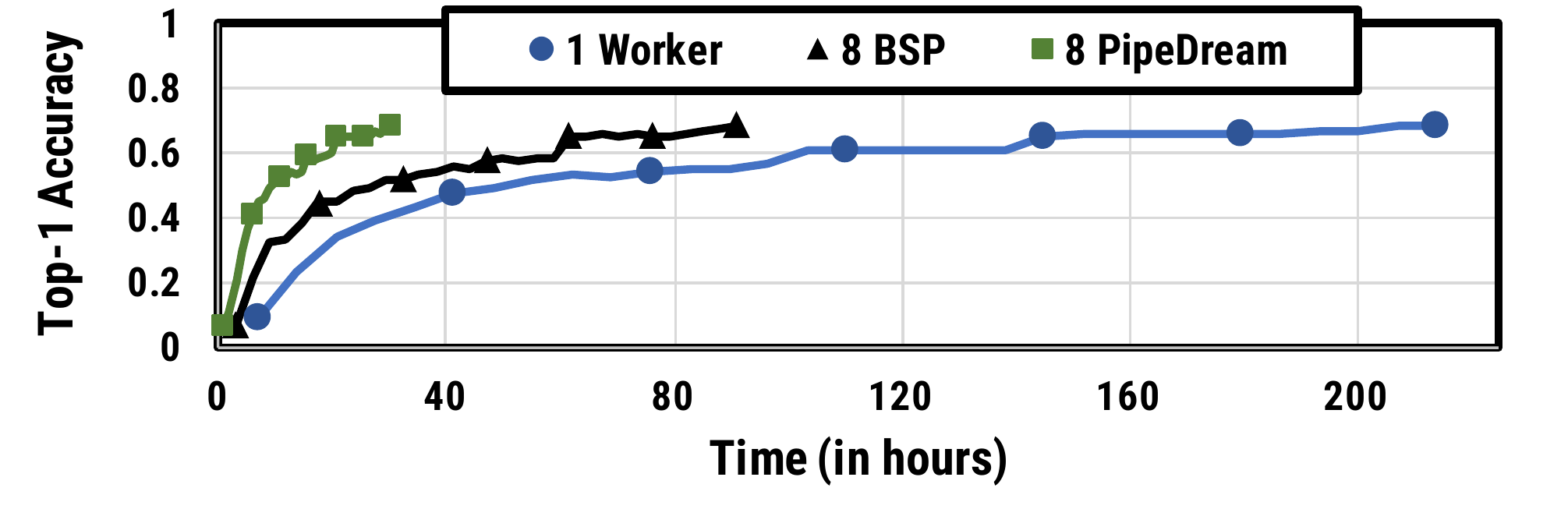}
    \caption{\vgg}
    \label{fig:cluster-a-8-vgg}
    \end{subfigure}
    \begin{subfigure}{0.45\textwidth}
    \includegraphics[keepaspectratio=1,width=1\columnwidth]{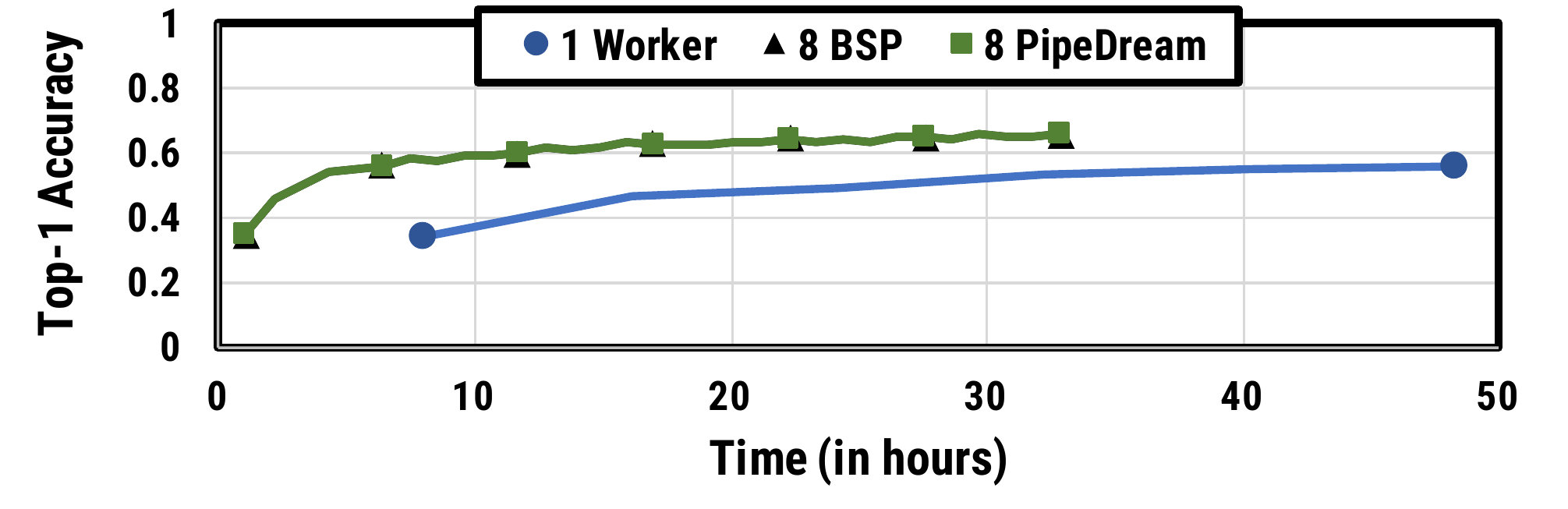}
    \caption{\inception}
    \label{fig:cluster-a-8-inception}
    \end{subfigure}
    \vspace{-0.1in}
    \caption{Accuracy vs. time for \vgg and \inception with 8 machines on \clusterA}
    \vspace{-0.15in}
    \label{fig:cluster-a-8}
\end{figure*}

\begin{figure*}[t]
    \centering
    \begin{subfigure}{0.45\textwidth}
    \includegraphics[keepaspectratio=1,width=1\columnwidth]{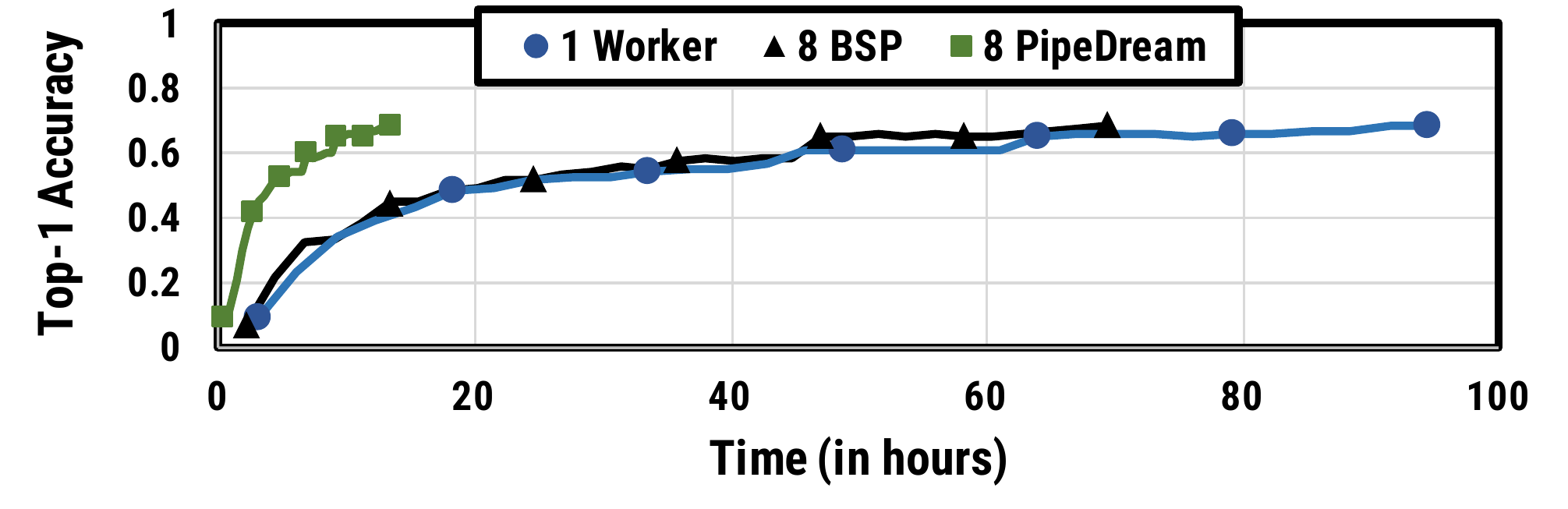}
    \caption{\vgg}
    \label{fig:cluster-b-8-vgg}
    \end{subfigure}
    \begin{subfigure}{0.45\textwidth}
    \includegraphics[keepaspectratio=1,width=1\columnwidth]{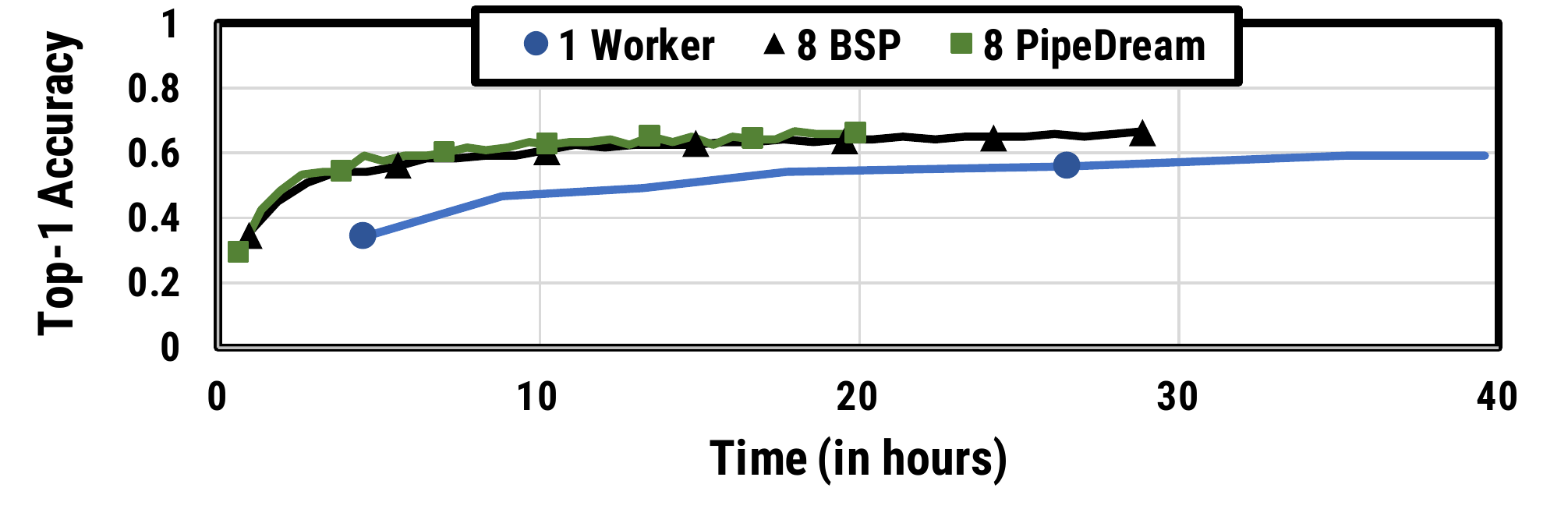}
    \caption{\inception}
    \label{fig:cluster-b-8-inception}
    \end{subfigure}
    \vspace{-0.1in}
    \caption{Accuracy vs. time for \vgg and \inception with 8 machines on \clusterB}
    \vspace{-0.15in}
    \label{fig:cluster-b-8}
\end{figure*}

\subsection{\system vs. Data Parallelism}

Table~\ref{table:eval-summary} summarizes results comparing \system with data-parallel training (BSP). For the three models, the table shows \system{}'s auto-generated configuration and the corresponding speedup in training time over single machine and data-parallel training (BSP). It also shows the communication reduction achieved by \system{} compared to data-parallel training. 

\textbf{\system Configurations.}
As described in Section~\ref{sec:layer-partitioning}, given a DNN model and a set of machines, \system{}'s optimizer selects the best configuration that partitions the layers of the model into multiple stages and assigns one or more machines to each stage. While most prior research has focused on improving data-parallel training, our results indicate that the best configurations are neither fully data-parallel nor fully model-parallel. 
In all but one of our experiments, the best \system{} configuration combines model parallelism, pipelining, and data parallelism; each of these configurations significantly outperform data-parallel training, thus highlighting the importance of pipeline parallelism.  \system's optimizer recommends data-parallel as the best configuration for \inception with 8 machines in \clusterA.

\textbf{Base Results: 8 Machines in \clusterA.}
Figure~\ref{fig:cluster-a-8} shows accuracy vs. training time for \vgg and \inception, using 8 machines in \clusterA, for both BSP and \system~\footnote{For Figure~\ref{fig:cluster-a-8}--~\ref{fig:cluster-a-4-16} each displayed point represents 5 epochs}. The first conclusion we draw is that for \vgg, BSP with 8 machines reduces training time by only $2.35\times$ compared to training with a single machine since with 8 machines, the communication overhead for \vgg in \clusterA is 72\%.
\system eliminates 95\% of this communication overhead thereby improving performance by $7.04\times$ compared to training with single machine ($2.99\times$ compared to BSP). Second, for \inception, the communication overhead with 8 machines on \clusterA is just 5\%. As a result, BSP achieves near perfect scaling with a $7.66\times$ speedup over a single machine. \system{}'s partitioning algorithm (Section ~\ref{sec:partition-algorithm}) selects a data-parallel configuration (no pipelining or model parallelism) to train Incpetion-v3 on \clusterA{}, thus matching the data-parallel BSP performance (Figure~\ref{fig:cluster-a-8-inception}).

\textbf{Effect of using faster compute (V100s).}
Figure~\ref{fig:cluster-b-8} shows accuracy vs. training time for \vgg
and \inception, using 8 machines in \clusterB, for both BSP and \system.
Compared to \clusterA, \clusterB employs faster V100 GPUs with 10Gbps interconnect
between the machines (as granted by the cloud provider). Thus, models
running on \clusterB have lower computation-to-communication ratios.
We note that the faster GPUs result in faster end-to-end training time
---e.g., training time for \vgg reduces from 220 hours on \clusterA to
little less than 100 hours on \clusterB (Figures~\ref{fig:cluster-a-8} and \ref{fig:cluster-b-8}).
We also observe that the higher communication overhead causes both BSP and \system
to scale less effectively to 8 machines. However, this increased communication
overhead affects BSP significantly more than \system.  Moving from \clusterA to
\clusterB, the speedup of \system over BSP increases from $2.99\times$ to
$5.12\times$ for \vgg. Even for \inception, \system improves training time
by 45\% compared to BSP on \clusterB.

Due to space constraints we do not present end-to-end training results for AlexNet~\cite{krizhevsky2012imagenet} and ResNet-50~\cite{he2016deep} models. Experiments with these models on \clusterB{} showed that \system{} provides a 1.21x and 6.78x throughput improvement for ResNet-50 and AlexNet respectively, compared to 8 machine data-parallel BSP.

\begin{figure}[t!]
    \centering
    \includegraphics[keepaspectratio=1,width=1\columnwidth]{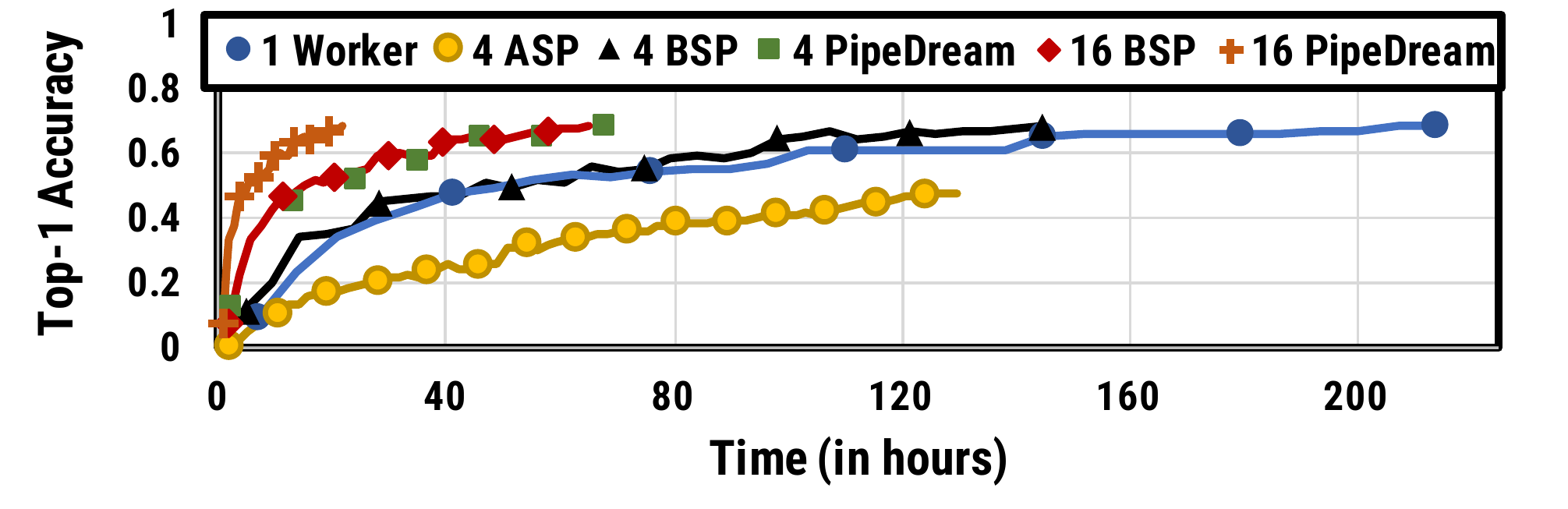}
    \caption{
        \label{fig:cluster-a-4-16}
        Accuracy vs. time for different configurations for VGG16 on \clusterA{} with 4 and 16 workers.
    }
    \vspace{-0.15in}
\end{figure}

\textbf{Effect of varying number of machines.}
With an increase in the number of machines, the communication overhead also increases for all models (as shown in Figure~\ref{fig:communication-ratio}). Figure~\ref{fig:cluster-a-4-16} compares the accuracy vs. training time for \vgg, using 4 and 16 machines in \clusterA, for BSP and \system. As expected, BSP scales poorly with increasing number of machines. With 4, 8, and 16 machines, BSP offers speedups of only $1.47\times$, $2.35\times$, and $3.28\times$, respectively, compared to single machine training. In contrast, \system offers speedups of $3.14\times$, $7.04\times$, and $9.86\times$ with 4, 8, and 16 machines compared to training on a single machine. Notably, \system with 4 machines is almost as good as BSP with 16 machines.

\textbf{Comparison to asynchronous parallel (ASP).}
To reduce communication overhead in data-parallel training, we experimented with running four machine data-parallel with ASP synchronization. Unlike BSP, which synchronizes the parameter data after every mini-batch, ASP has no synchronization, and the workers use the most recent parameter data \emph{available}. Figure~\ref{fig:cluster-a-4-16} also shows the accuracy vs. time curve for ASP with 4 machines in \clusterA. Due to ASP's poor statistical efficiency, \system{} reaches a 48\% accuracy 7.4x faster than ASP data-parallel, even though ASP has no communication overhead.

\textbf{Training a Recurrent Neural Network.}
VGG16 and Inception-v3 are Convolutional Neural Networks (CNNs), used for tasks such as image classification.  We also evaluate \system{}'s performance on the S2VT model, which is a sequence-to-sequence recurrent neural network that generates descriptive captions for videos. For many recurrent neural networks consisting of LSTM layers, BSP data-parallel training scales poorly. For S2VT, BSP with 4 machines reduces training time by only 1.1x compared to single machine training. This is because with 4 machines, the communication overhead for S2VT on \clusterA{} is 70\%. \system reduces the communication overhead by 95\% compared to BSP, in turn slashing training time by $3.34\times$ compared to single machine training ($3.01\times$ compared to BSP).

\subsection{Value of Data Parallelism in \stages{}}
\label{sec:compare-pipelining}

Figure~\ref{fig:pipeline-vs-pipedream-graph} plots the reduction in training time compared to single machine training for three parallelization approaches: 1)~simple model parallelism (no pipelining or data parallelism), 2)~ pipeline parallelism without data parallelism (no \stage replication), and 3)~\system (combined model parallelism, pipelining and data parallelism). The figure shows these results for training \vgg using 4 and 8 machines on \clusterA. 

\begin{figure}[h]
    \centering
    \includegraphics[width=\columnwidth]{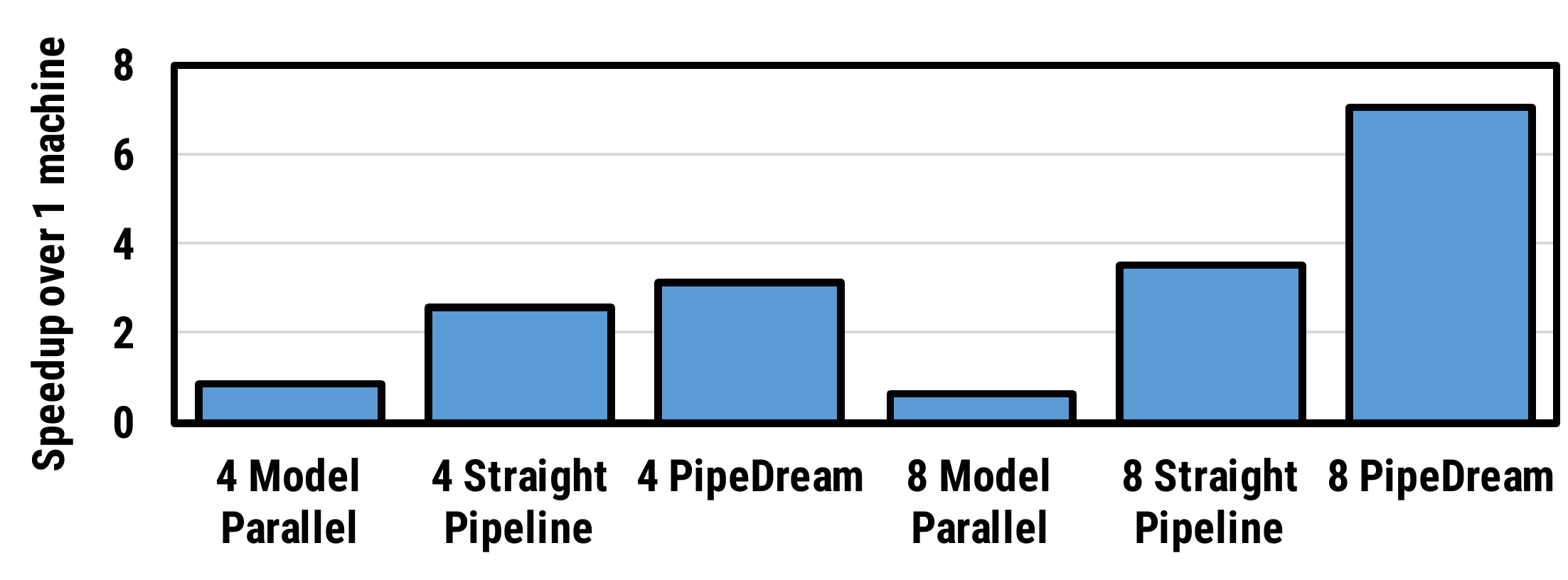}
    \caption{Model Parallelism vs. Pipeline Parallelism vs. \system{} for VGG16 on \clusterA{}}
    \label{fig:pipeline-vs-pipedream-graph}
\end{figure}

\textbf{Model Parallelism.} Simple model parallelism uses only one machine at any point in time, and hence is slower than the single machine configurations. We also implement model-parallel configurations in Tensorflow~\cite{tensorflow}, which does not support pipelining but implements model parallelism as described in this paper; we observe similar performance drop-offs. 

\textbf{Straight Pipelines.}
Combining model parallelism with pipelining results in straight pipeline configurations (no data parallelism compared to \system{}). Straight pipeline configurations greatly reduce training time compared to single machine training---$2.56\times$ and $3.49\times$ with 4 and 8 machines, respectively. In fact, these improvements are better than data parallel training, which achieves corresponding speedups of $1.47\times$ and $2.35\times$ compared to single machine training.

\textbf{Pipeline Parallelism.}
\system{}'s pipeline parallelism provides the biggest reductions in training time---$3.14\times$ and $7.04\times$ with 4 and 8 machines compared to single machine training. These results demonstrate that a combination of pipelining, model parallelism and data parallelism achieve faster training than either model parallelism, model parallelism with pipelining, or data parallelism. 

\section{Conclusion}
\label{sec:concl}

Pipeline-parallel DNN training addresses the communication overheads that bottleneck data-parallel training of very large DNNs.
\system{} automatically partitions and aggressively pipelines DNN training across worker machines.
Compared to state-of-the-art approaches, \system{} is up to 5$\times$ faster in ``time to target accuracy'' for experiments with five different DNNs on two different clusters.


\setlength{\bibsep}{2pt plus 1pt}  
\small
\bibliography{bib}
\bibliographystyle{abbrvnat}
}{
}


\end{document}